\documentclass[]{aa}
\usepackage{graphicx}
\usepackage{txfonts}
\usepackage{natbib}
\def\idm#1{{\mbox{\scriptsize #1}}}
\newcommand\Chi{{(\chi^2_\nu)^{1/2}}}
\newcommand\stara{{HD~208487}}
\newcommand\starb{{HD~190360}}
\newcommand\starc{{HD~188015}}
\newcommand\stard{{HD~114729}}
\newcommand\stare{{HD~147513}}

\begin{document}
   \title{About putative Neptune-like extrasolar planetary candidates}
   \author{Krzysztof Go\'zdziewski \and Cezary Migaszewski}
   \offprints{k.gozdziewski@astri.uni.torun.pl}
   \institute{Toru\'n Centre for Astronomy, PL-87-100 Toru\'n, Poland}
   \date{Received; accepted}
   \abstract{
   We re-analyze the precision radial velocity (RV) data of \stara, \starb, 
   \starc{} and \stard{}. These stars are supposed to
   host Jovian companions in long-period orbits.
   }
   {
   We test a hypothesis that the residuals of the 1-planet model of the RV or 
   an irregular scatter of the measurements about the synthetic RV curve may be
   explained by the existence of additional planets in short-period orbits.
   }
   {
   We perform a global search for the best fits in the orbital parameters space
   with genetic algorithms and simplex method. This makes it possible to verify
   and extend the results  obtained with an application of commonly used FFT-based
   periodogram analysis for identifying the leading periods.
   }
   {
   Our analysis confirms the presence of a periodic component in the RV of
   \starb{} which may correspond to a hot-Neptune planet. We found four new
   cases when the 2-planet model yields significantly better fits to the RV data
   than the best 1-planet solutions. If the  periodic variability of the
   residuals of single-planet fits has indeed  a planetary origin then
   hot-Neptune planets may exist in these extrasolar systems. We estimate their
   orbital periods in the range of 7--20~d and minimal masses  about of 20
   masses od the Earth.
   }
   {}
   \keywords{extrasolar planets---Doppler technique---
   \stara{}---\starb{}---\starc{}---\stard{}---\stare{}}
   \authorrunning{Go\'zdziewski \& Migaszewski}
   \titlerunning{New Neptun-like extrasolar candidates?}
   \maketitle
%
\section{Introduction}
%
The precision radial velocity (RV) surveys for extrasolar planets approach the
limit of detection of Neptune-like companions. In the past year, 14-Earth mass
planet has been announced by \cite{Santos2004}. A low-mass companion has been
found  in the ${\rho}^{1}$~Cancri system~\citep{McArthur2005}. Almost at the
same time, a Neptune-like planet has been also discovered by \cite{Butler2004}.
Very recently, \cite{Rivera2005} announced a discovery of $\sim 7$ Earth-mass
planet in $\sim 2$d orbit about the famous Gliese~876 planetary system.  Also
\cite{Vogt2005} announced a new multi-planet system about HD~190360 involving a
Neptune-mass planet in $\simeq 17.1$d orbit with a Jupiter-like companion with
$\sim 3000$d orbital period.  These discoveries change the view of multi-planet
systems. The found low-mass planets likely are rocky bodies rather than massive
gaseous hot-Jupiters. The architecture of their environment, involving also
giant planets in relatively distant orbits, begin to resemble our own Solar
system.

There are already speculations as to which of the extrasolar systems  involving
Jupiter-like planets can contain smaller companions. At present, such objects
cannot be clearly detected because of small amplitude of the 
contributed RV signal
and not frequent enough observations.  Nevertheless, the existence
of such planets seems to be natural. The dynamical relaxation of planetary
systems \citep{Adams2003}  can lead to final  configurations with orbits filling
up the dynamically  available space of stable orbits~\citep{Laskar2000}. In
particular,  many  extrasolar systems  containing  distant Jupiter-like planets
may contain smaller, still undetectable companions, which  can easily survive 
in  a close neighborhood of the parent star.  \cite{Baraffe2005} show that the
three recently detected hot-Neptune planets (about GJ~436, $\rho^{1}$~Cnc,
$\mu$~Arae) may originate from more massive gas giants which have undergone
significant evaporation. This work suggests that hot-Neptunes and hot-Jupiters
may share the same origin and evolution history and  they can be as frequent as
are hot-Jupiters. The numerical simulations  by \cite{Brunini2005} show that the
formation of compact systems of Neptune-like planets close to solar type stars
could be a common by-product of planetary accretion. In turn, these theoretical
works are  supported by several new detections of hot-Neptune and hot-Saturn
candidates
\cite[e.g.,][]{Bonfils2005,Udry2005,LoCurto2005,Fischer2005}\footnote{ For
frequent updates of the references, please see the Extrasolar Planets
Encyclopedia by Jean Schneider, http://www.obspm.fr/planets.}.

The detection of Neptune-mass exoplanets is a breakthrough in the field but the
inevitable problem still lies in the limitations of the Doppler technique. The
radial velocity signal of small, Neptune-like planets revolving close to the
host star has the semi-amplitude of a few~meters per second, i.e., at a similar
range as the measurement errors and chromospheric variability of the RV. This
can be a source of erroneous interpretation of the observations.  For instance,
in a recent work, \cite{Wisdom2005b} claims that the orbital period of 2.38d of
companion~d  detected by \cite{McArthur2005} in ${\rho}^{1}$~Cancri system is an
alias of two periods originating from the orbital motion of planets~b and~c and,
instead, he found an evidence of an another planet in $\sim 261$~d orbit. 
Actually, a proper interpretation of the observations is still very difficult. 
Often, they can be equally well modeled by qualitatively different
configurations and even the number of planets cannot be resolved without doubt.
An example will be given in this paper too.

In this work, for a reference and a test of our approach, we  re-analyze the
published RV data of two stars:  \stara{} \citep{Tinney2005}--- 
\cite{Gregory2005a} claims that a second planet accompanies a confirmed Jovian
companion~b, and \starb{} \citep{Vogt2005}---the discovery team  claims the
existence of a hot-Neptune and a Jovian planet in long-period orbit.  Next,
using the same method, we study other three extrasolar systems:  \starc{}
\citep{Marcy2005}, \stard{} \citep{Butler2003} and \stare{} \citep{Mayor2004},
in which Jupiter-like companions have been discovered in long-period orbits.
However, the residuals of the 1-Keplerian fits are either relatively large or we
see "suspicious" scatter of the residuals. We use the available measurements to
test a hypothesis that additional short-period planets exist in these systems
and their signals may be already hidden in the existing data, but likely at the
detection limit. Our reasoning follows the theoretical predictions that
Neptune-like planets can easily form and survive close to the parent star. The
main difficulty is a small number of measurements, typically about of 30--40
data points spread over several years. It makes it  very difficult to determine
orbits of  such  planetary candidates without a doubt. In particular, this
concerns  the identification of the orbital periods with the application of 
somehow traditional Fourier-based analysis. Instead, we prefer {\em a global}
search for the best-fit 2-planet configurations using the  genetic algorithms
and scanning $\Chi$ in the  multidimensional space of orbital parameters. This
could be an alternative to the already classic Lomb-Scargle periodogram
\cite[e.g.,][]{Butler2004}  or sophisticated Bayesian-based analysis, which is
recently proposed by \cite{Gregory2005a,Gregory2005b}. That author claims an 
evidence of additional planets using the same RV measurements which are modeled
by 1-planet configurations by the discovery teams, about HD~73526 and HD~208487.

\section{The fitting algorithm, numerical setup and tests}
%
In all cases studied in this paper, we assume that the second, putative planet
is closer to the star than the already known companion~b in relatively
long-period orbit. To model the RV signal, we use the standard formulae
\cite[e.g.,][]{Smart1949}. For every planet, the contribution to the reflex
motion of the star at time $t$ is the following:
\begin{equation}
 V_{\idm{r}}(t) = K [ \cos (\omega+\nu(t))  + e \cos \omega] + V_0,
\label{eq:eq1} 
\end{equation}
where $K$ is the semi-amplitude, $\omega$ is the argument of pericenter,
$\nu(t)$ is the true anomaly involving implicit dependence on the orbital period
$P$ and the time of periastron passage $T_{\idm{p}}$, $e$ is the eccentricity,
$V_0$ is the velocity offset. There are some arguments that is best to interpret
the derived fit parameters $(K,P,e,\omega,T_{\idm{p}})$ in terms of  Keplerian
elements and minimal masses related to Jacobi coordinates
\citep{Lee2003,Gozdziewski2003e}.  

To search for the best-fit solutions, we applied a kind of hybrid optimization.
A single program run starts the genetic algorithms  (GAs) optimization. In
particular,  we use the PIKAIA code by \cite{Charbonneau1995}. The GAs has many
advantages over more popular gradient-type algorithms, for instance, the
Levenberg-Marquardt scheme \citep{Press1992}. The power of GAs lies in their
global nature, the requirement of knowing only the $\Chi$ function and an ease
of constrained optimization. In particular, GAs permit to define parameter
bounds or add a penalty term to $\Chi$
\citep{Gozdziewski2003e,Gozdziewski2005a}.

In our code, the initial population of randomly chosen 2048 members represents
potential solutions to the RV model. The code produce 128 new off-spring
generations. The best fit found by PIKAIA is refined by the simplex algorithm of
Melder and Nead~\citep{Press1992}. In that method, we set the fractional
tolerance  for $\Chi$ variability to $10^{-15}$. Such procedure runs thousands
of times for every analyzed data set. Thanks to non-gradient character of both
algorithms we may bound the space of examined parameters, according to specific
requirements.  Usually,  due to a small number of  measurements, the best-fit 
$e_{\idm{c}}$  is barely constrained and tends to be large, in particular in the
range of very short periods. Nevertheless,  we assumed that the tidal
circularization limits the eccentricity of the short-period orbit and we
arbitrarily  set its upper bound to 0.3. The orbital periods of putative planets
are searched in the range of [2,136]~d.

The internal errors of the RV data are rescaled according to 
$
 \sigma^2 = \sigma_{\idm{m}}^2 + \sigma_{\idm{j}}^2,
$ 
where $\sigma$ is the joint uncertainty,  $\sigma_{\idm{m}}$ and
$\sigma_{\idm{j}}$ is the internal error and adopted variance of stellar jitter,
respectively. Typically, we choose  $\sigma_{\idm{j}}$ following estimates for
Sun-like dwarfs \citep{Wright2005} or we set its value as quoted by the discovery
teams. The particular estimates are given in Table~1.

The best fits  found in the entire  hybrid search  are illustrated by 
projections onto planes of particular parameters of the RV model, Eq.~1.  
Usually, we choose the $(P,K)$- and $(P,e)$-planes. Marking the elements within
the formal $1\sigma,2\sigma,3\sigma$ confidence intervals of the best-fit
solution  (which is always marked in the maps by two crossing lines) we have a
convenient way of  visualizing the shape of the local minima of $\Chi$ and to
obtain reasonable estimates of the fit errors \citep{Bevington2003}.

Still, even we find very good best-fit configurations, a problem persists: is
the  obtained fit not an artifact caused by specific time-distribution of
measurements or the stellar jitter?  In order to answer for this question we
apply the test of scrambled velocities \citep{Butler2004}. We remove the
synthetic  RV contribution of the firstly confirmed planet from the RV data, 
according to its best-fit parameters.   Next, the residuals are randomly
scrambled, keeping the exact moments of observations, and  we search for the
best-fit elements of the second (usually, inner) planet.  For  uncorrelated
residual signal we should get close to the Gaussian distribution of $\Chi$. For
every synthetic data set, the minimization is  performed with the hybrid method.
Let us note, that in this case we deal with 1-planet model (and 6 parameters
only) thus the large population in the GA step and precise simplex code assure
that the best-fit solution can be reliably found  in one single run of the
hybrid code. The runs of the code are repeated  at least 20,000 times for  every
analyzed data set. If the best  fit corresponding to the real (not scrambled)
data lies on the  edge of this distribution, such experiment assure us that the
real data produce much smaller $\Chi$ (and an rms) than the most frequent
outcomes obtained for the scrambled data. Such a test  provides a reliable
indication that the residuals are not randomly distributed. A likelihood
$p_{\idm{H}}$ that the residuals are only a white noise can be measured by  the
ratio of the best fits with $\Chi$ less or equal to that one  corresponding to the
real data against all possible solutions. 

Unfortunately, even with a positive outcome of the test of scrambled residuals
we still are not done. Even if we find a strong indication that the residuals
are not randomly distributed, and we find a periodic component in the residual
signal,  then we still cannot be sure that it has a planetary origin and it does
not eliminate jitter as a source of the RV signal.  We are warned by the referee 
(and here we literally follow his arguments) that jitter can have some spectral
power usually associated with the rotation period of the star, but other peaks
can be present. The spectral structure of very low-amplitude jitter is not
well-known yet, but for large amplitude jitter in active star, the power
spectrum can show several peaks and aliases. One example is HD~128311 in
\cite{Butler2003}; an another one is, possibly, \stare{} \citep{Mayor2004}  analyzed in this
work. Therefore, the noisy periodic signal may be due to jitter and it  may be
possible that some secondary  periods, which are present in the residual
signals, are corresponding to aliases of the stellar rotation period reinforced
by the noise. This is the main reason why our work cannot be considered in terms
of a discovery paper; additional measurements and analysis are required to
confirm (or withdraw)  the hypothesis about the planetary nature
of the RV variability. Our basic goal
is to test the approach on a few non-trivial and, possibly, representative
examples. We try to find characteristics of the RV data sets which can confirm
the periodogram-based approach \cite[e.g.,][]{Butler2004} and, possibly, may be
helpful to extend its results. Our analysis may be useful to speed-up the
detections of new planets and planning the strategy of future observations.

\subsection{Test case I: \stara}
%
The first test case of \stara{} \citep{Tinney2005} illustrates  nuances of
modeling the RV data. The observational window  of this star spans about of 
2500~d but the number of data points is only~31. The discovery team found a
planetary companion with minimal mass of 0.7~$m_{\idm{J}}$ and orbital period
about of 130~d.  An rms of this 1-planet solution is $\sim 7$~m/s, while the
joint uncertainty is about of $6$~m/s, assuming jitter estimate of 3~m/s, a
reasonable value for the quiet, 6~Gyr old star with $R_{HK}'\simeq
-4.90$~\citep{Tinney2005}. The data have been reanalyzed by \cite{Gregory2005a}.
He applied his Bayes-periodogram method and found much better 2-planet solution,
explaining the residual signal of the 1-planet solution by the presence of  a
distant, second companion in 998~d orbit. This solution has significantly
better rms of 4.2~m/s and $\Chi\simeq 0.83$ than the best single planet
solution, $\Chi=1.27$.

\begin{figure*}
   \hspace*{0.1cm}\centerline{\hbox{\includegraphics[]{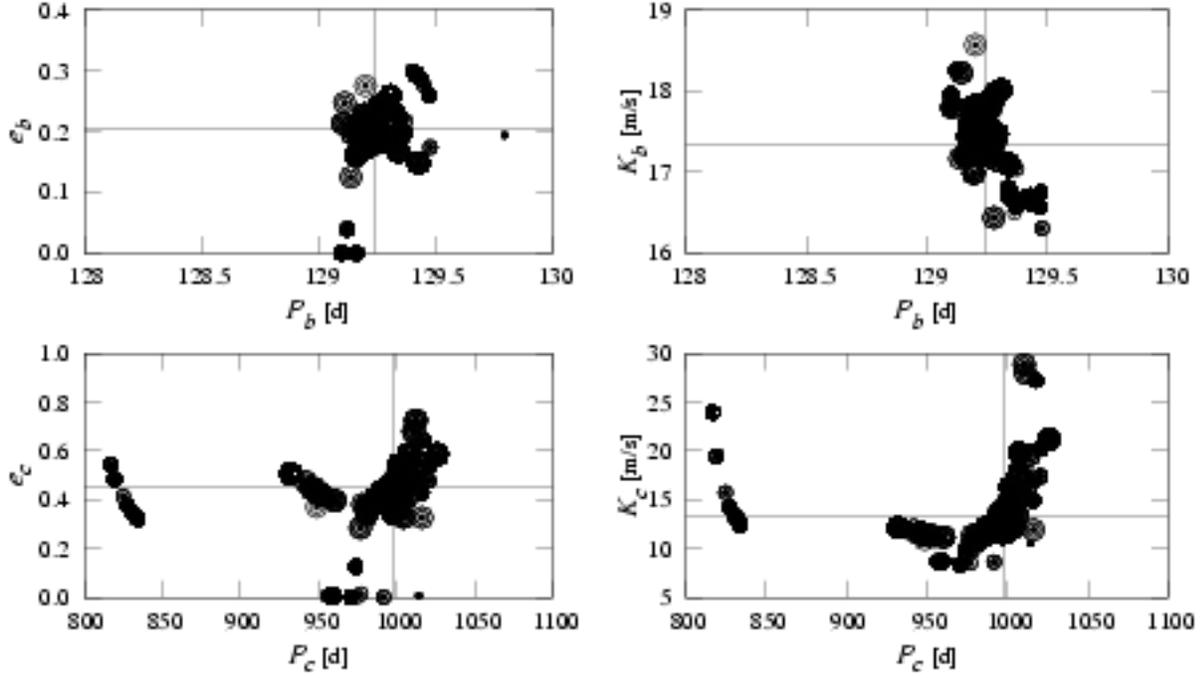}}}    
   \caption{
   The parameters of the best-fit solutions to the 2-planet model of the
   \stara{} RV data projected onto the ($P,e$)- and ($P,K$)-plane. In this
   search,  a more distant companion to the formerly known Jupiter-like body in
   $\simeq 130$~d orbit is assumed. Upper plots are for the inner planet, bottom
   plots are for the outer companion. The values of  $\Chi$ of the best-fit
   solutions are marked by the size of symbols. Largest circle is for $\Chi$
   equal to 0.619 ; smaller symbols are for 1$\sigma$ solutions
   with $\Chi \in [0.619,0.710)$, 
   2$\sigma$ solutions
   with $\Chi \in [0.710,0.841)$ and
   3$\sigma$ solutions
   with $\Chi \in [0.841,1.0)$ (smallest, filled circles), respectively.   
}
\label{fig:fig1}
\end{figure*}

The analysis of the \stara{} data is a good reference for our simpler
approach. First we did the hybrid search for the best 2-planet configurations,
assuming that  130~d $< P_{\idm{c}} <$ 1200~d.  The results of this search are
illustrated in Fig.~\ref{fig:fig1}. Curiously, the best-fit has $\Chi\simeq
0.62$ and an rms $\simeq 3.1$~m/s, so it is  apparently better than the fit
quoted by \cite{Gregory2005a}. The  best-fit parameters
$(K,P,e,\omega,T_{\idm{p}}-T_0)$ are:  $(17.532~\mbox{m/s}, 129.224~\mbox{d},
0.195, 188^{\circ}.232, 1421.112~\mbox{d})$ and $(13.117~\mbox{m/s},
997.953~\mbox{d}, 0.433, 190^{\circ}.470, 2358.258~\mbox{d})$ for the inner and
outer planet, respectively (where $T_0$=JD~2,450,000) and $V_0=9.64$~m/s. Note
much  larger eccentricity of the outer planet than in the fit by
\cite{Gregory2005a}, $e_{\idm{c}}\simeq 0.2$. The $1\sigma$ scatter of the
derived solutions around the best-fit (Fig.~\ref{fig:fig1}) provides  the error
estimate of the orbital elements. In general, the parameters of the inner planet
are well fixed, while the elements of the outer body have much larger formal
uncertainties. The errors of $K_{\idm{c}},P_{\idm{c}},e_{\idm{c}}$ are about of
100~d, 0.3, and $10$~m/s, respectively.

\begin{figure*}
   \hspace*{0.1cm}\centerline{\hbox{\includegraphics[]{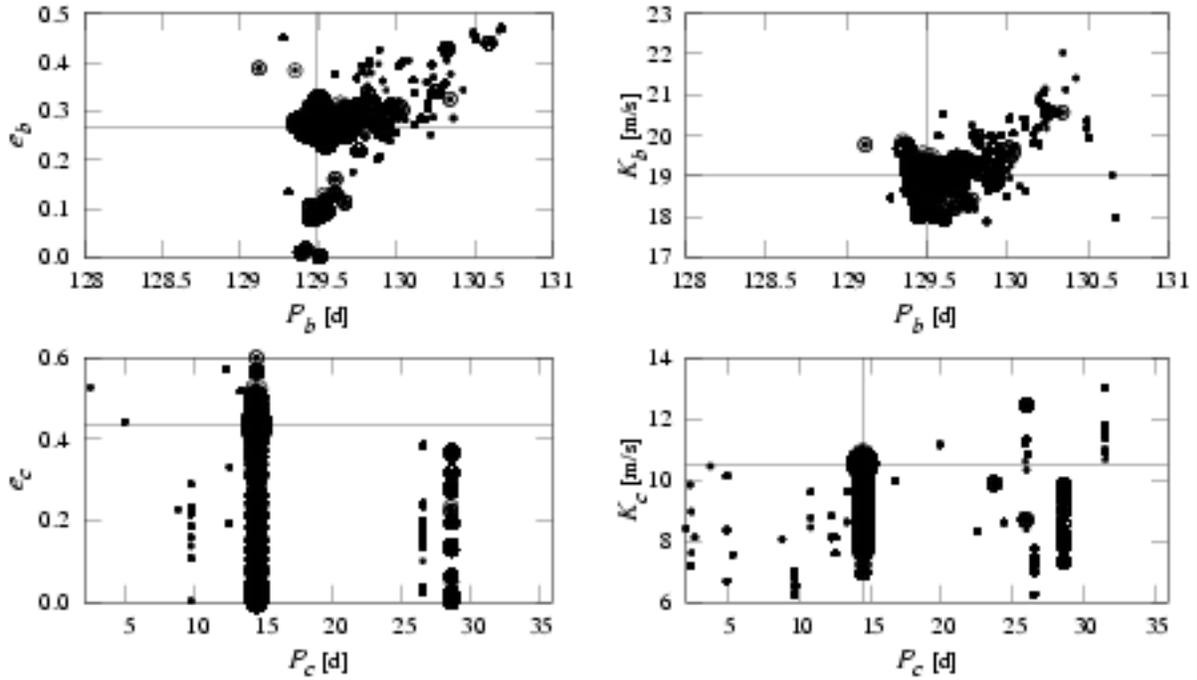}}}
   \caption{
   The parameters of the best-fit solutions to the 2-planet model of the
   \stara{} RV data  projected onto ($P,e$)- and ($P,K$)-plane. In this
   search,  an inner companion to the formerly known Jovian planet in $\simeq
   130$~d orbit is assumed.   Upper plots are for the outer planet, bottom plots
   are for the inner companion. The values of  $\Chi$ of the best-fit solutions
   are marked by the size of symbols. Largest circle is for $\Chi=0.581$ (the
   best-fit solution, Table~1); smaller symbols are for solutions within
   $1\sigma$ interval of the best-fit  with $\Chi \in [0.581,0.677)$, 
   2$\sigma$ solutions
   with $\Chi \in [0.677,0.813)$ and
   3$\sigma$ solutions
   with $\Chi \in [0.813,0.98)$ (smallest, filled circles), respectively.   
}
\label{fig:fig2}
\end{figure*}

\begin{figure}
   \centering
   \hbox{\includegraphics[]{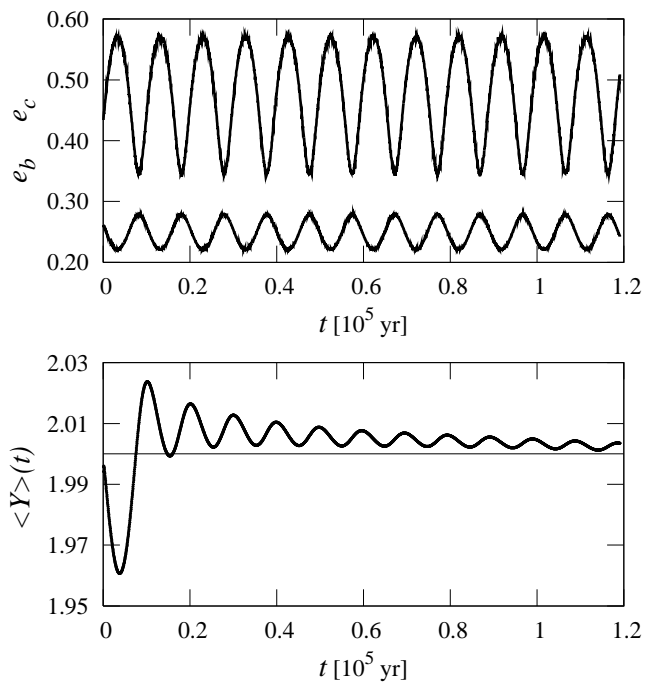}}
   \caption{
   Stability analysis of the orbital fit with  putative Neptune-like body in the
   \stara{} system. The upper panel is for the  eccentricities, the lower panel
   is for MEGNO computed for $3\cdot 10^5$ orbital periods of the outer
   companion.
}
\label{fig:fig3}%
\end{figure}

Next, we explored the inner space to the $\sim 130$~d orbit, assuming  2~d $<
P_{\idm{c}} <$ 130~d. Surprisingly, we found ever better fit, given in Table~1,
with $\Chi\simeq 0.581$ and an rms $\simeq 2.82$~m/s. The orbital elements  of
the best-fit solutions found in the entire search are projected onto the  
parameter planes  and they are illustrated in Fig.~\ref{fig:fig2}. Note a
small scatter of the elements of the outer planet and two sharp minima in
$P_{\idm{c}}$.   The best-fit  has $P_{\idm{c}} \simeq 14.5$~d and 
$e_{\idm{c}}\simeq 0.42$.  Such large eccentricity may contradict the
requirement of a tidally circularized orbit. However, the  orbital evolution of
the whole system reveals large-amplitude variations of $e_{\idm{c}}$
(Fig.~\ref{fig:fig3}) which are forced by the mutual interactions between
companions.  We also face an another trouble, a small periastron-apoastron
distance of the two putative planets $\sim 0.2$~AU. Nevertheless, the system
appears to be dynamically stable. The evolution of  the MEGNO indicator
\citep{Gozdziewski2001a} computed over $3\cdot 10^5 P_{\idm{b}}$ is illustrated
in the lower panel of Fig.~\ref{fig:fig3}. Its oscillations about of~2 mean a
quasi-periodic configuration and strictly confirm apparently regular behavior 
of $e_{\idm{b,c}}$.

\begin{figure}
   \centering
   \hbox{\includegraphics[width=3.5in]{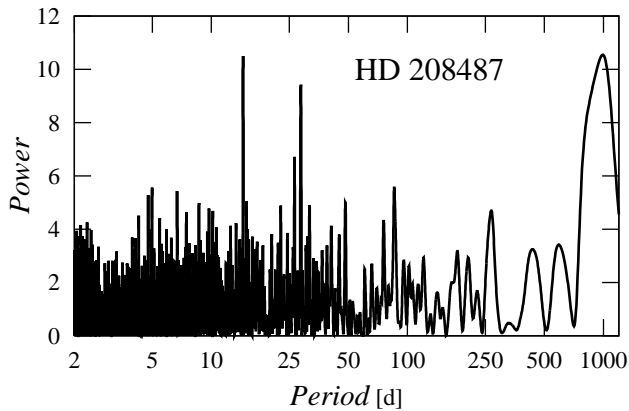}}
   \caption{Lomb-Scargle periodogram of the residuals of 
   the Jupiter-like companion's  with $P_{\idm{b}} \sim 130$~d
   about \stara{}.
   }
\label{fig:fig4}%
\end{figure}

\begin{figure*}
   \centering
   \hbox{
   \hbox{\includegraphics[]{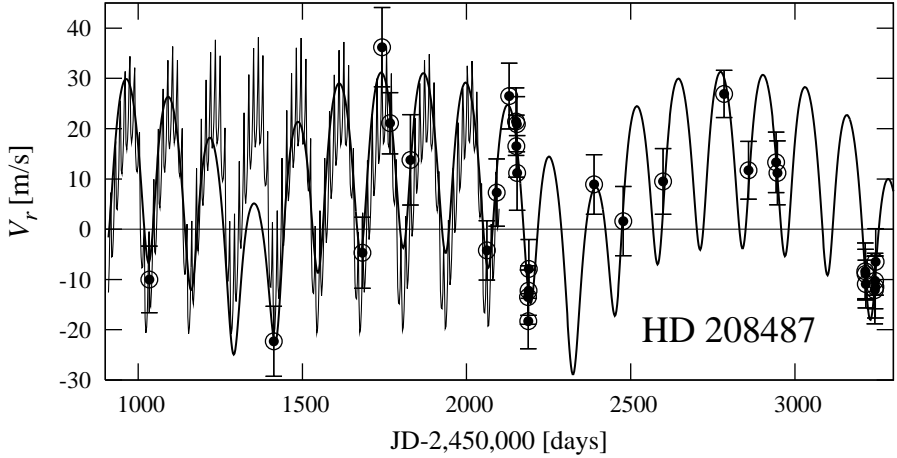}}
   }
   \caption{
   Synthetic RV signals of the two solutions to the RV data of \stara.  The
   thick line is for the outer companion to the known $130$~d planet.  The thin
   line is for the putative inner companion. The original RV data are marked
   with error-bars. For a clear illustration, only a part of the second curve is
   plotted.
}
\label{fig:fig5}%
\end{figure*}

For a reference, we computed the Lomb-Scargle (LS) periodogram 
(Fig.~\ref{fig:fig4}) of the residuals of the  dominant 1-planet signal  having
the period of $130$~d.  The largest peaks about of 28~d and 998~d  (the period
of a hypothetic long-period planet) most possibly are aliasing with the best-fit
period $\simeq 14.5$~d.  However, this  period is favored by the smallest
$\Chi$.  The aliasing of the two dominant periods is illustrated in
Fig.~\ref{fig:fig5}. This figure shows the synthetic RV signals of the putative
configurations with the short- and long-period companions of planet~b.
Curiously, the orbital periods of the two planets in the model with the
short-period orbit are very close to the 9:1 ratio. This again could be an
effect of aliasing, nevertheless the best-fit solution with single planet  has
$P_{\idm{b}}$ about of $14.5$~d and much larger rms $ \sim 7$~m/s than in the best
2-planet solution.

\begin{figure}
   \centering
   \hbox{\includegraphics[width=3.5in]{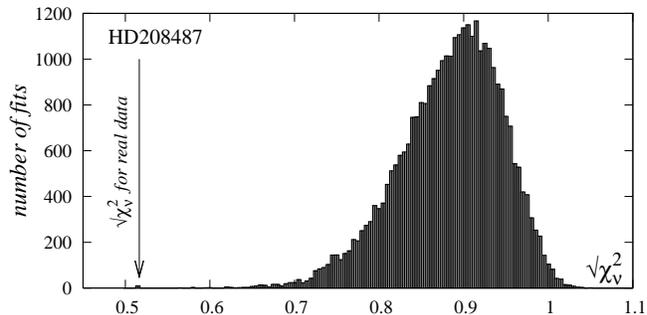}}
   \caption{
   A histogram of $\Chi$ for Keplerian fits to 33,000 sets of scrambled RV
   residuals of the synthetic signal of the Jupiter-like planet in $\sim
   130$~d orbit  (\stara, see Table~1). The position of the best-fit 
   to the real data is marked with an arrow.
   }
\label{fig:fig6}%
\end{figure}

Some more interesting results brings the test of scrambled residuals,
Fig.~\ref{fig:fig6}. It appears that the real  residual signal lies far from the
maximum of the Gaussian-like distribution and it is unlikely that the residuals
are only a white noise ($p_{\idm{H}}\simeq  3\cdot 10^{-5}$). Also
$K_{\idm{b}}\simeq 10.5$~m/s is much larger than the joint uncertainty of the
measurements. Because \stara{} is a very old and quiet G-dwarf with low
chromospheric activity \citep{Tinney2005},  the hypothesis of a hot sub-Neptune
planet in this extrasolar  system seems to be very appealing. A peculiarity of
this system could be the proximity to the 9:1 mean motion resonance of its
companions.

\subsection{Test case II: \starb{}}
%
In a recent paper, \cite{Vogt2005} published the precision RV data for \starb{}.
The variability of the RV is explained by  the existence of two planetary
companions: a Jupiter-like planet in long-period orbit, $P_{\idm{b}} \sim
3000$~d, and a Neptune-like body in $\simeq 17.1$~d orbit. The authors conclude
that the existence of the smaller planet is the best explanation of the residual
signal  of the single planet solution. Its  contribution has  small
semi-amplitude $K\simeq 4.6$~m/s. Yet the orbital period is about of a half of
the rotational period of the star, 36--44~d, and the RV variability may be an
effect of spot complexes on the stellar  surface \citep{Vogt2005}.

\begin{figure*}
   \hspace*{0.1cm}\centerline{\hbox{\includegraphics[]{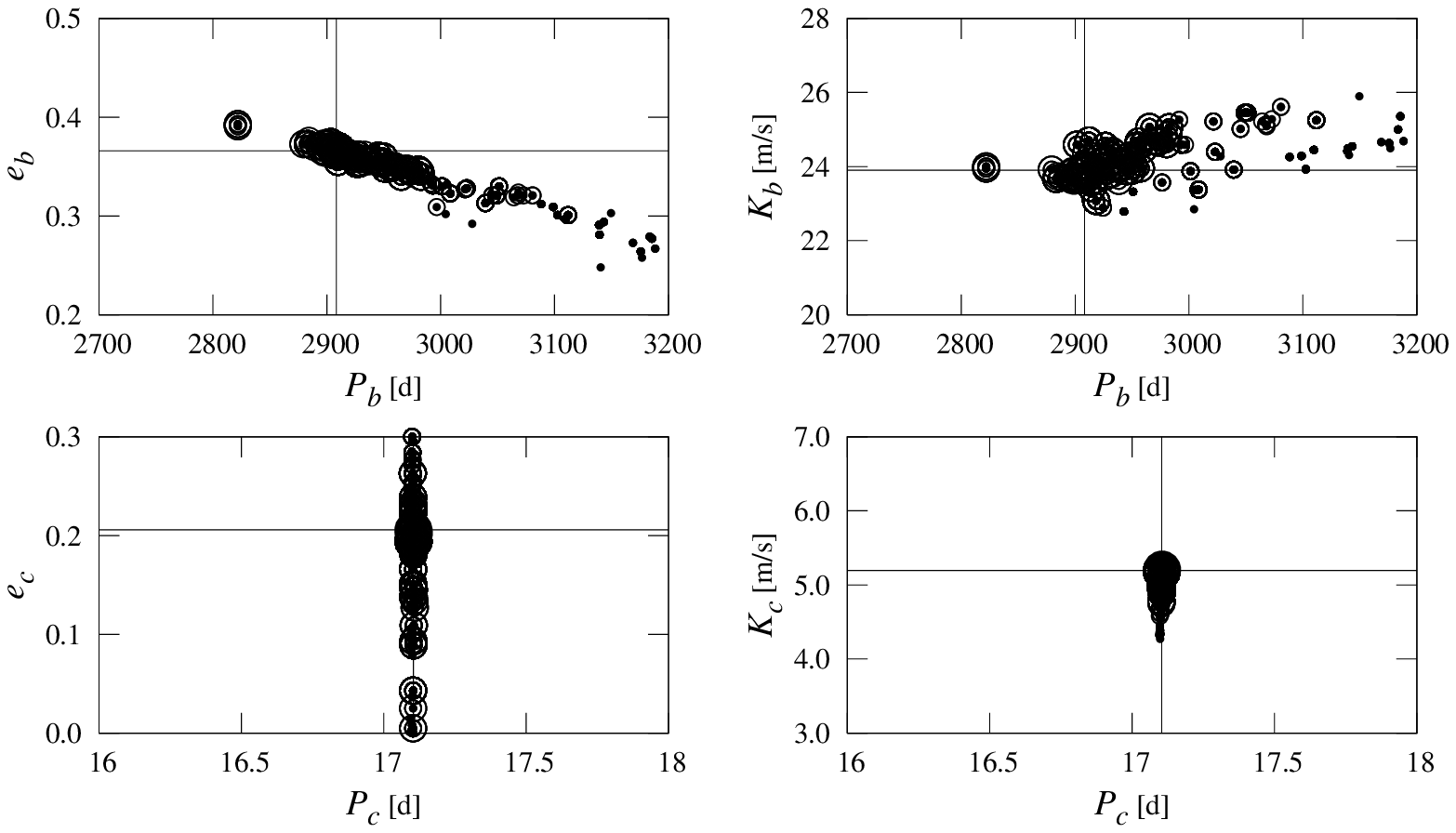}}}
   \caption{
   The parameters of the best-fit solutions to the 2-planet model of the
   \starb{} RV data  projected onto the ($P,e$)- and ($P,K$)-plane.  Upper
   plots are for the outer planet, bottom plots are for the inner companion. The
   values of  $\Chi$ of the best-fit solutions are marked by the size of
   symbols. Largest circle is for  $\Chi\simeq0.679$  (the best-fit solution,
   Table~1); smaller symbols are for 1$\sigma$ solutions with $\Chi \in
   [0.679,0.70)$, 
   2$\sigma$ solutions
   with $\Chi \in [0.70,0.74)$ and
   3$\sigma$ solutions
   with $\Chi \in [0.74,0.79)$ (smallest, filled circles), respectively.   
}
\label{fig:fig7}
\end{figure*}

The elements of a few hundred fits gathered by the hybrid search are
illustrated in Fig.~\ref{fig:fig7} (the elements
of the best-fit are given in Table~1). The
period of the inner planet corresponds to a  single, very deep minimum of
$\Chi$. It is clear, that $e_{\idm{c}}$ has a large error $\sim 0.15$. Our
fit is slightly better than that one quoted by \cite{Vogt2005} but $e_{\idm{c}}\simeq
0.2$ is also different as well as we have got slightly larger $K_{\idm{c}}
\simeq 5.2$~m/s. Note that we  obtain a similar uncertainties of 100~d for
$K_{\idm{b}}$ and of 0.05 for $e_{\idm{b}}$. 

\begin{figure}
   \centering
   \hbox{\includegraphics[]{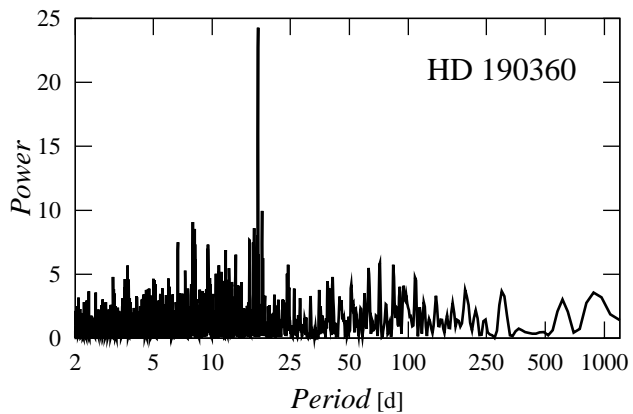}}
   \caption{
   Lomb-Scargle periodogram of the residuals of  the long-period signal which is
   seen in the \starb{} data.
}
\label{fig:fig8}%
\end{figure}

The Lomb-Scargle periodogram (Fig.~\ref{fig:fig8}) of the residuals of the
long-period orbit reveals a strong, dominant peak at $17.1$~d. This an excellent
confirmation of the results of the hybrid search. The test of scrambled
residuals reveals that the residual signal is hardly random. The
maximum of $\Chi$ for scrambled data is shifted by $\simeq 0.4$ from the value
of $\Chi$  for the real data.

\begin{figure}
   \centering
   \hbox{\includegraphics[width=3.5in]{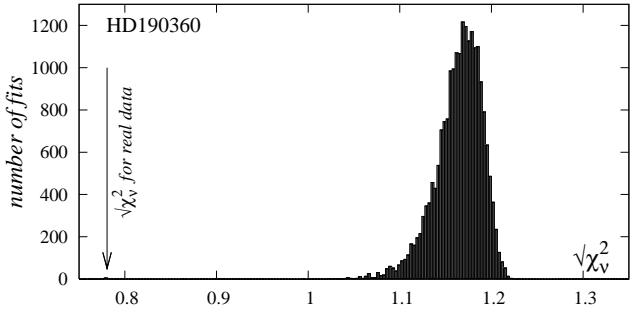}}
   \caption{
   A histogram of $\Chi$ for Keplerian fits to 20,000 sets of scrambled RV
   residuals of the synthetic signal of the outer planet in \starb{} (see
   Table~1). The position of the best-fit to the real data is marked with an
   arrow.
   }
\label{fig:fig9}%
\end{figure}

Now, having so clear picture,  we analyze a smaller sample of measurements. To
perform this test, we selected  38~first measurements spanning  $\sim 2200$~d.
The number of data points and their time-coverage are similar to other cases
which we study in this work.  The described above procedure was repeated on such
synthetic data set.

\begin{figure*}
   \hspace*{0.1cm}\centerline{
         \hbox{\includegraphics[]{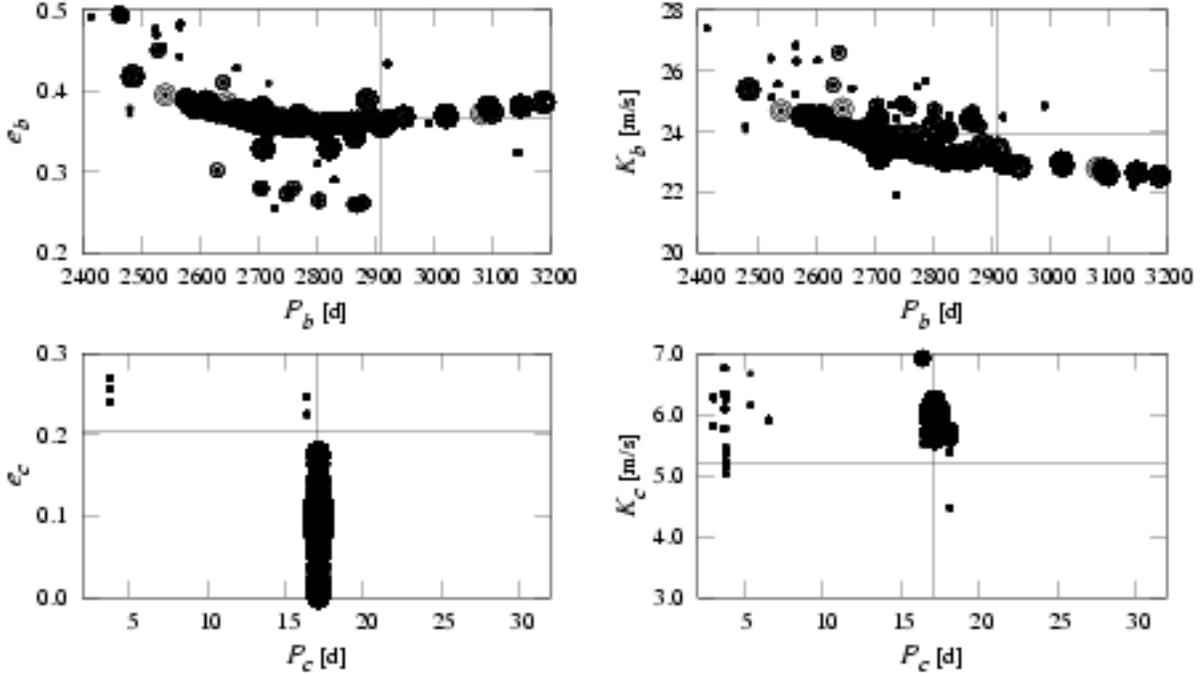}}}
   \caption{
   The parameters of the best-fit solutions to the 2-planet model of the  RV
   data of \starb{} limited to the first 38 measurements, projected onto
   ($P,e$)- and ($P,K$)-plane.  Upper plots are for the outer planet, bottom
   plots are for the inner companion. The values of  $\Chi$ of the best-fit
   solutions are marked by the size of symbols. Largest circle is for
   $\Chi\simeq 0.733$;  smaller symbols are for
   1$\sigma$ solutions
   with $\Chi \in [0.733,0.793)$, 
   2$\sigma$ solutions
   with $\Chi \in [0.793,0.89$ and
   3$\sigma$ solutions
   with $\Chi \in [0.89,1.00)$ (smallest, filled circles), respectively.
   Vertical and horizontal lines mark
the best-fit obtained for the full data (see Fig.~\ref{fig:fig7}
 and Table~1).
}
\label{fig:fig10}
\end{figure*}

\begin{figure}
   \centering
   \hbox{\includegraphics[]{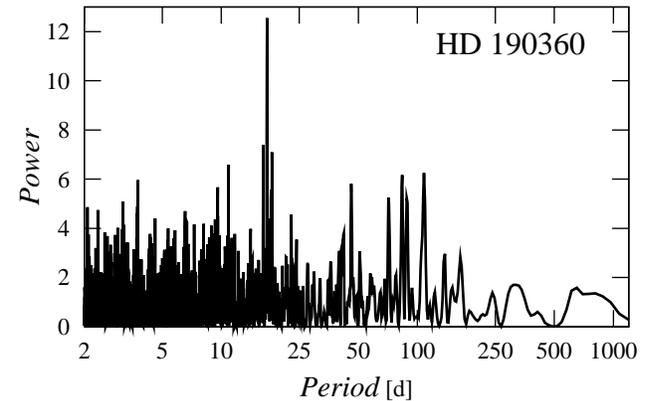}}
   \caption{
   Lomb-Scargle periodogram of the residuals of the long-period
   signal in \starb{} when only 38 first measurements are analyzed.
   Compare with Fig.~\ref{fig:fig8} for the full data set.
}
\label{fig:fig11}%
\end{figure}

\begin{figure}
   \centering
   \hbox{
   \hbox{\includegraphics[width=3.5in]{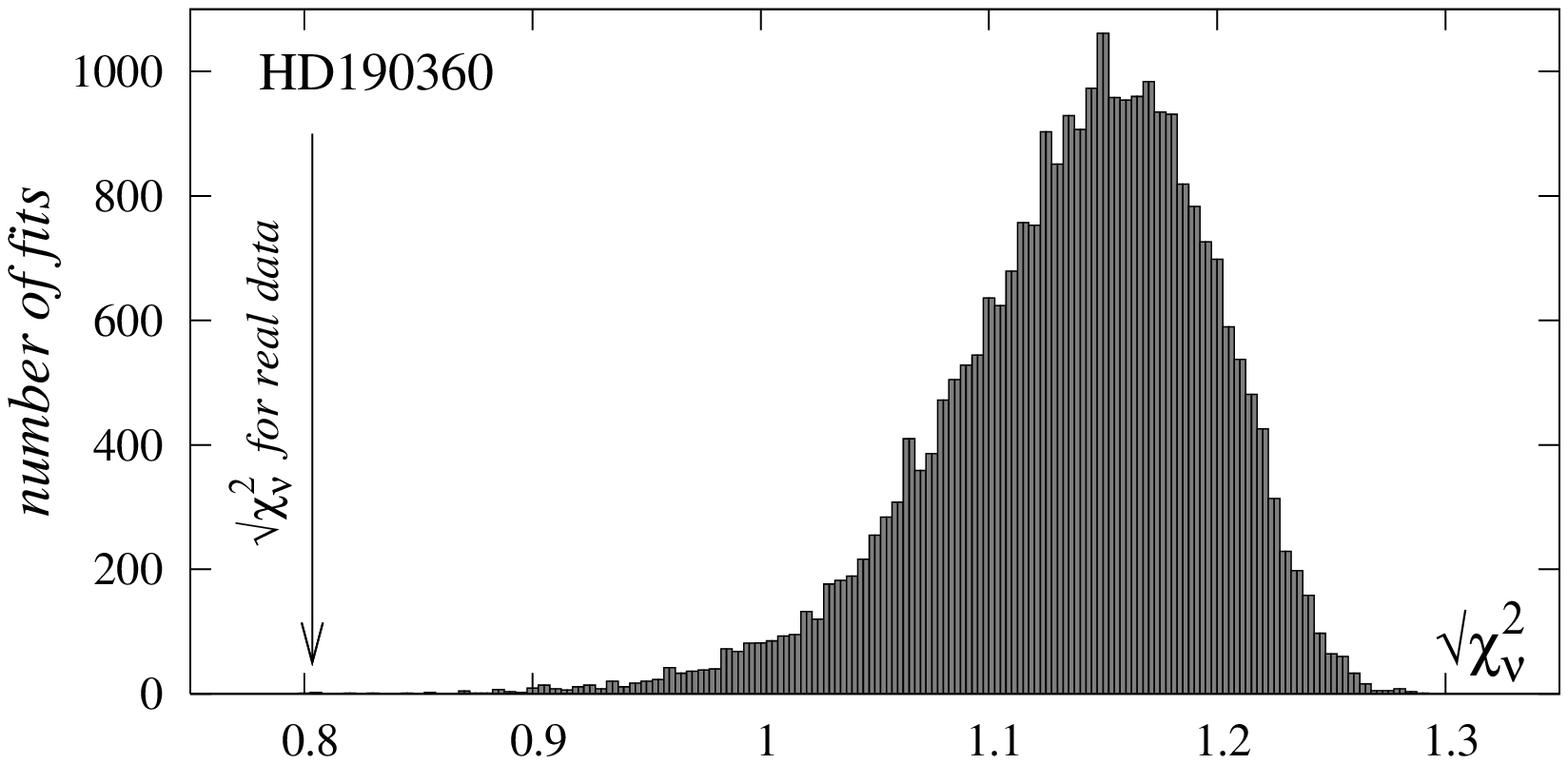}}
   }
   \caption{
   A histogram of $\Chi$ for Keplerian fits to $\sim 27,000$ sets of scrambled RV
   residuals of the synthetic signal of the outer planet in \starb{} 
   for the first 38~measurements. The position of the best-fit  for the
   real data is marked with an arrow. Compare with Fig.~\ref{fig:fig9}. }
\label{fig:fig12}%
\end{figure}

Still, the best-fit period of the inner planet can be clearly identified
(Fig.~\ref{fig:fig10}) but the semi-amplitude $K_{\idm{c}}$ has been slightly
shifted. The elements of the outer planet are  determined with much larger
errors than the uncertainties  obtained for the full data. In particular, the
error of $P_{\idm{b}}$ is about of 350~d. The dominant period is still well
recognizable in the Lomb-Scargle periodogram, Fig.~\ref{fig:fig11}. Finally, the
test of scrambled residuals also give a clear picture, albeit the histogram
(Fig.~\ref{fig:fig12}) has much larger width and  $\Chi$ of the real data is 
closer to the maximum of the distribution.

Even for a half of the available measurements, the periodic nature of the residual
signal of the single-planet fit can be clearly recognized. One might conclude
that the inner planet  around  \starb{} might be detected about 2--3~years ago.

The results obtained for the \starb{} data provide  a valuable test of our
method and a good reference point to further analysis. Let us recall, that the
model with two planets yields low residuals  with an rms of $3.5$~m/s and $\Chi$
much smaller than~1, similarly to the three systems which  are described below.

\section{\starc{}}
\begin{figure*}
   \centering
   \hbox{
      \hbox{\includegraphics[width=3.4in]{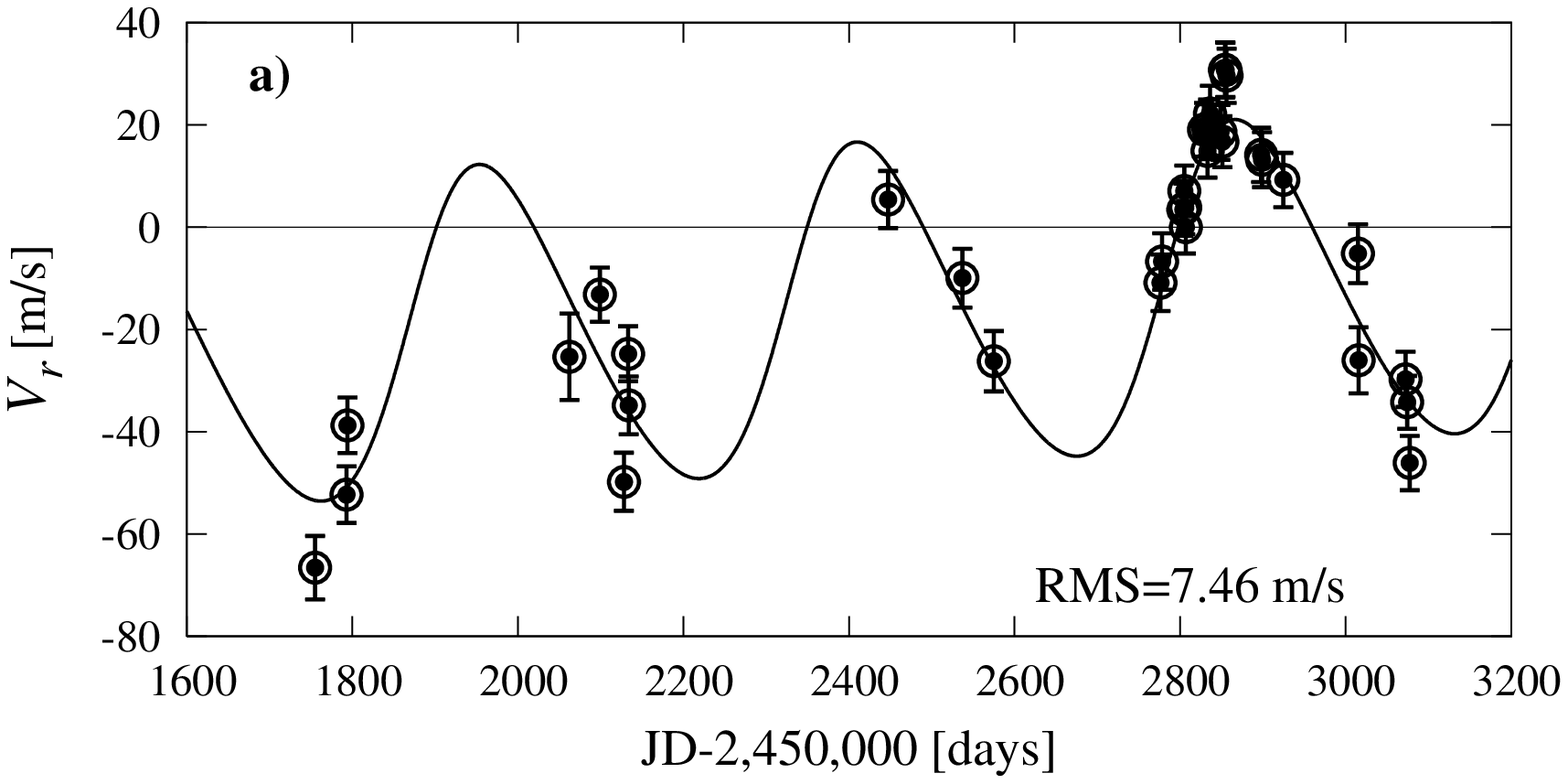}}
      \hbox{\includegraphics[width=3.4in]{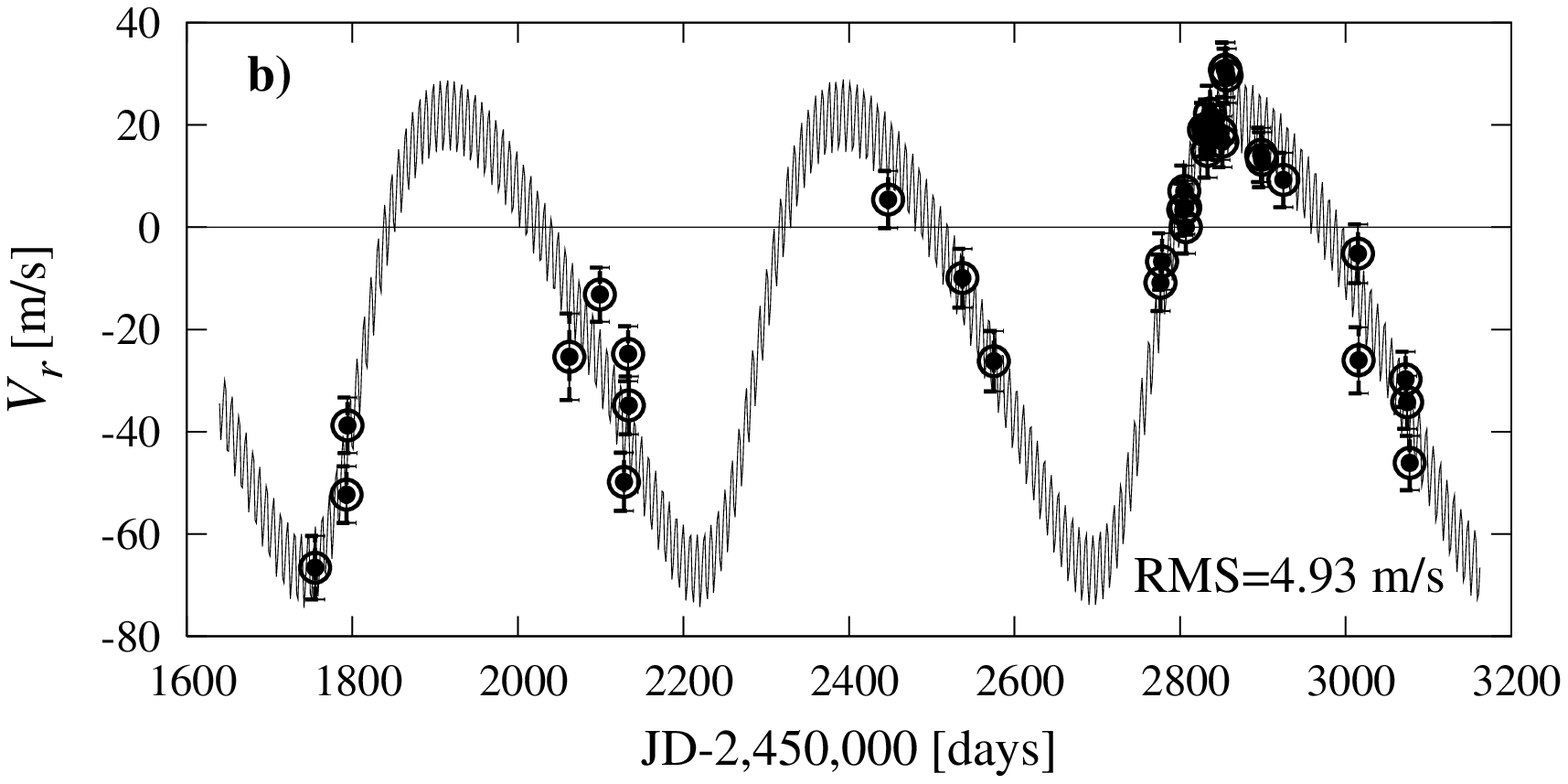}}
   }
   \hbox{
      \hbox{\includegraphics[width=3.4in]{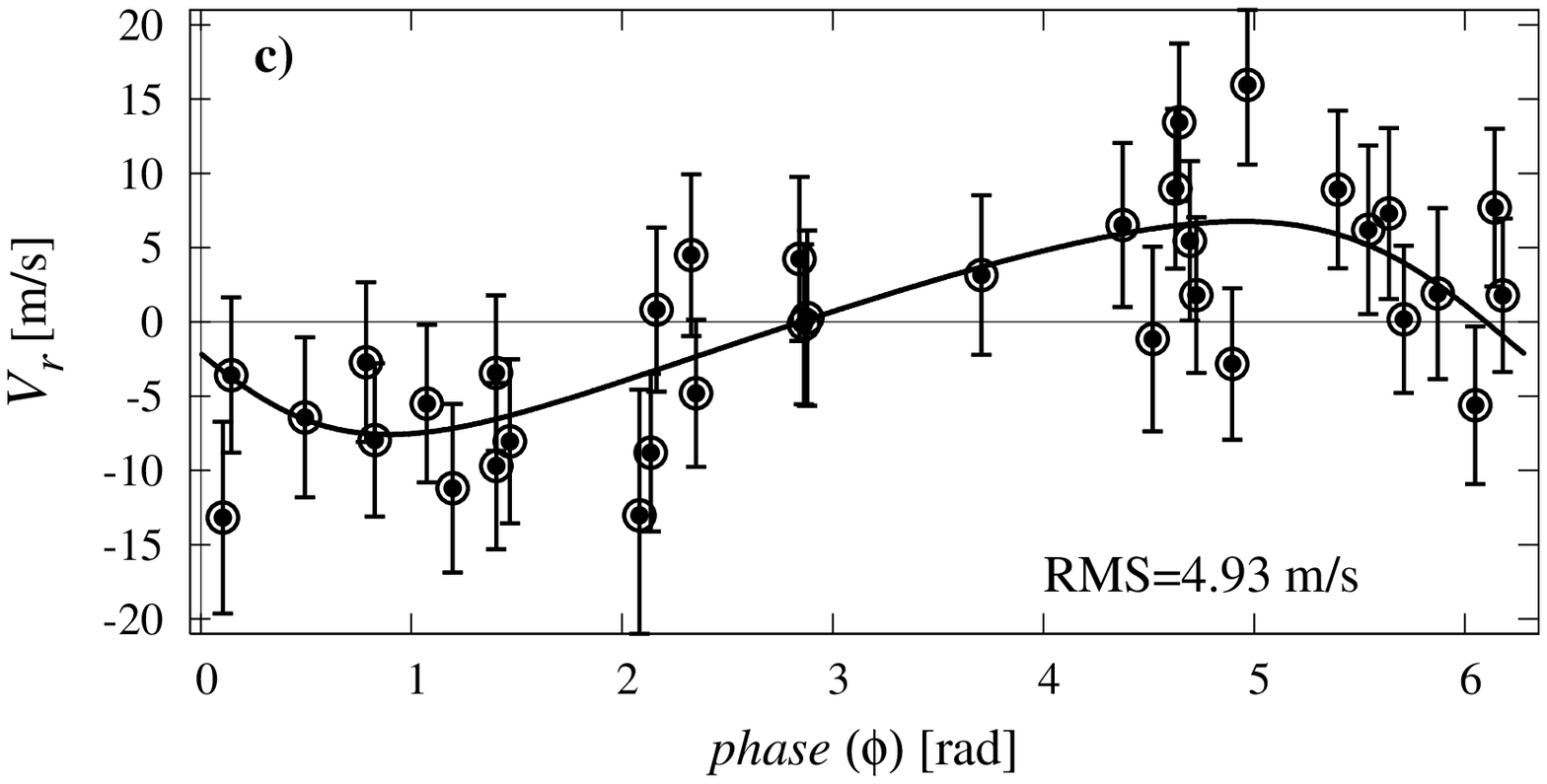}}
      \hbox{\includegraphics[width=3.4in]{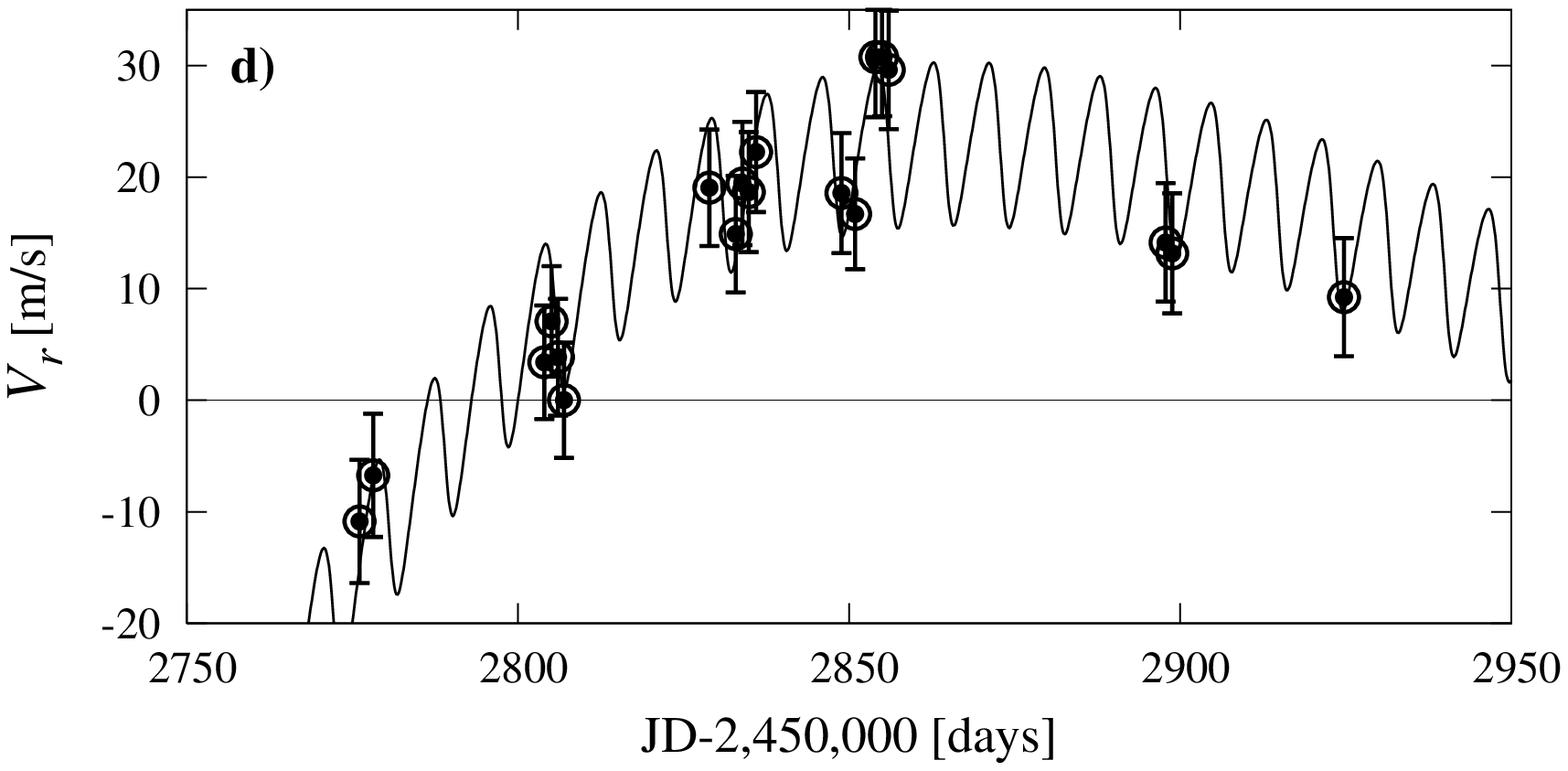}}
   }
   \caption{
    The best-fit solutions to the RV data published by \cite{Marcy2005} for
    \starc{}. {\bf a}) 1-Kepler model with a linear trend which is apparently
    present in the data.  {\bf b}) 2-Kepler model is consistent with the
    measurement error,  $\Chi \simeq 1.02$, rms$\simeq 4.9$~m/s.  {\bf c})
    Period-phased  RV signal of the inner companion. {\bf d}) A close-up of the
    synthetic signal of the best-fit 2-planet configuration (see Table~1).
   }
\label{fig:fig13}%
\end{figure*}

%
Recently, \cite{Marcy2005} published the precision RV of \starc{}. The
variability of the RV is explained by  the presence of  a planetary companion
with minimal mass 1.26~$m_{\idm{J}}$ and orbital period $\sim 1.25$~yr.  Our
attention has  been drawn on  large rms of this 1-planet solution, $\simeq
8.6$~m/s. The mean error  of the RV data is about of 4.1~m/s and expected 
stellar jitter $\simeq 4$~m/s \citep{Marcy2005}. The mean measurement error
added in quadrature to  the jitter is about of 5~m/s, still much smaller than
the rms of 1-planet solution.  However, this can be  partially explained by a
linear trend apparently present  in the RV signal. If one adds such a trend to
the fit model, the rms goes down by 1~m/s \citep[][ also our
Fig.~\ref{fig:fig13}a]{Marcy2005}.  The trends in the RV data  frequently 
indicate  a distant body \cite[e.g.,][]{Marcy2005a,Jones2002a,Gozdziewski2003e}.
Nevertheless, we assumed that the second putative body is not a distant
companion, but rather a planet closer to the star than already detected
companion~b.

The elements of  the best-fit solution found in the entire hybrid search are
illustrated in Fig.~\ref{fig:fig13} (panels b,c,d) and its  elements are given
in Table~1. The synthetic Doppler signal is plotted over the data points in
Fig.~\ref{fig:fig13}b. Surprisingly, the  apparent slope of the RV data  can be
only an artifact related to a specific time-distribution of measurements.  The
value of $\Chi \simeq 1.021$ indicates a statistically good solution. Its rms is
about of 5~m/s, consistent with the adopted estimate of the mean uncertainty.
The semi-amplitude of the inner planets' signal is  $\sim 7$~m/s. Assuming the
mass of the host star $\simeq 1.08$~M$_{\sun}$ \citep{Marcy2005},  we derived its
minimal mass about of 0.07~$m_{\idm{J}}$ and semi-major axis about of 0.08~AU.  
Thus, the new  putative object can be classified as a hot-Neptune planet. 

\begin{figure*}
   \hspace*{0.1cm}\centerline{\hbox{\includegraphics[]{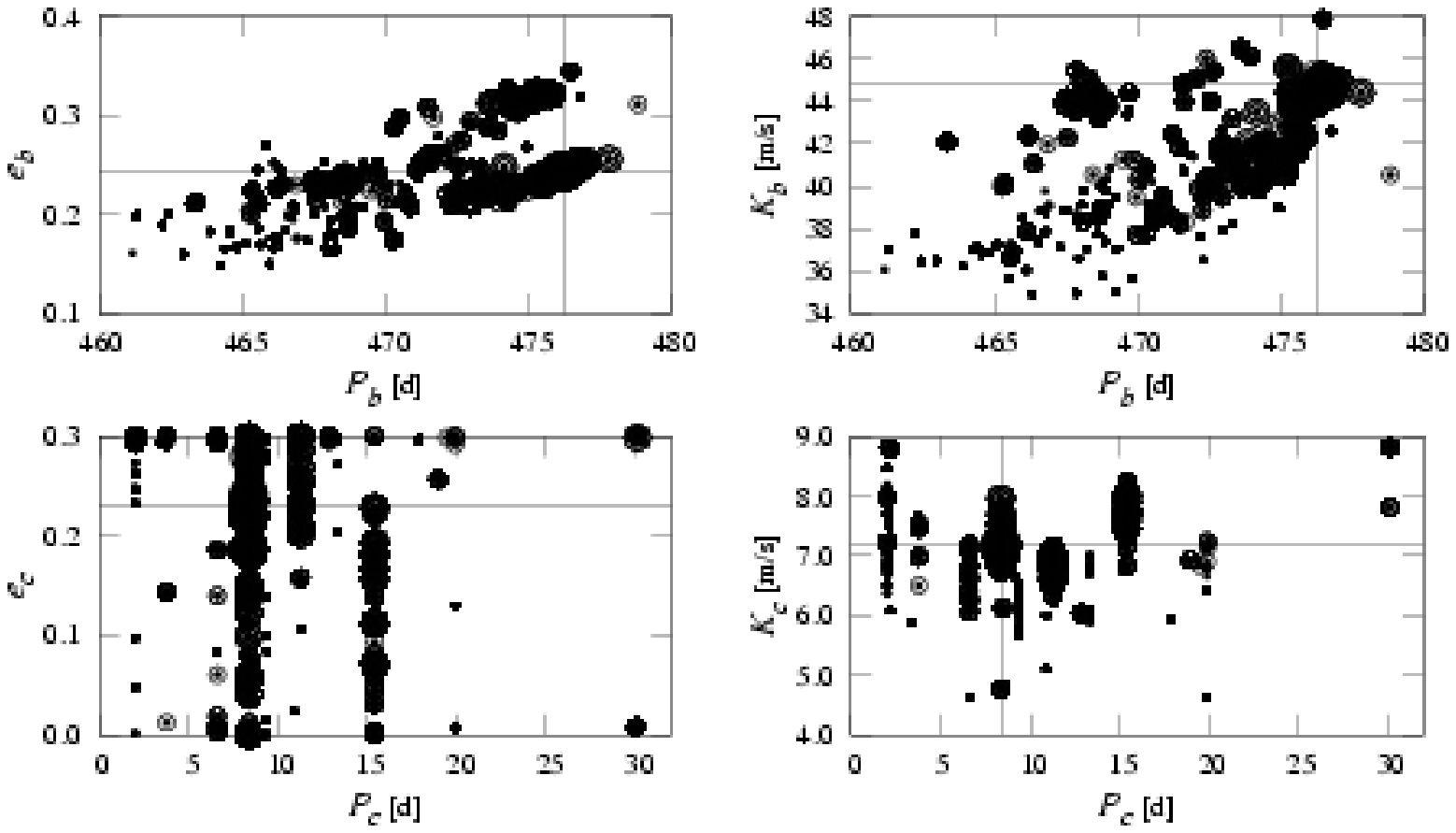}}}
   \caption{
   The elements of the best-fit solutions to the 2-planet model of the \starc{} 
   RV data projected onto the ($P,e$)- and ($P,K$)-plane.  Upper plots are for
   the outer planet, bottom plots are for the inner companion. The $\Chi$ of the
   best-fit solutions are marked by the size of symbols. Largest circle is for
   $\Chi=1.021$ (the best-fit solution, Table~1); smaller symbols are for
   1$\sigma$ solutions with $\Chi \in [1.02,1.07)$, 
   2$\sigma$ solutions
   with $\Chi \in [1.07,1.14)$ and
   3$\sigma$ solutions
   with $\Chi \in [1.14,1.25)$ (smallest, filled circles), respectively.
}
\label{fig:fig14}%
\end{figure*}

According to the $1\sigma$-level of $\Chi$, marked in Fig.~\ref{fig:fig14}, the
uncertainty of $e_{\idm{b}}$ is $\sim 0.1$, and of $K_{\idm{b}}$ is $\sim
5$~m/s.  Similarly, the $1\sigma$ uncertainties of the elements of  the inner
planet can be estimated by $\sim 0.3$ and $1$~m/s, respectively. There are
visible  several local minima in the $(P_{\idm{b}},e_{\idm{b}})$-plane, which
are very narrow with respect to $P_{\idm{c}}$, about of 0.05~d at the $1\sigma$
level. It is  clear that the eccentricity  of the inner companion (also its
argument of periastron --- not shown here) cannot be well fixed.  

A local minimum at  $P_{\idm{c}} \simeq 15.37$~d which  also well models the
observations (at the $1\sigma$ confidence interval of the best-fit solution) may
correspond to an alias of  the best-fit period 8.386~d. It could be also an
alias of the rotational period $\sim 30$~d of the star, assuming that it is
smaller than the 36~d estimate by \citep{Marcy2005}. Indeed, there is   a
minimum of $\Chi$ at the $1\sigma$-level of the best-fit about of  $30$~d and
$e_{\idm{c}}\sim 0.3$, see Fig.~\ref{fig:fig14}. 

\begin{figure}
   \centering
   \hbox{\includegraphics[]{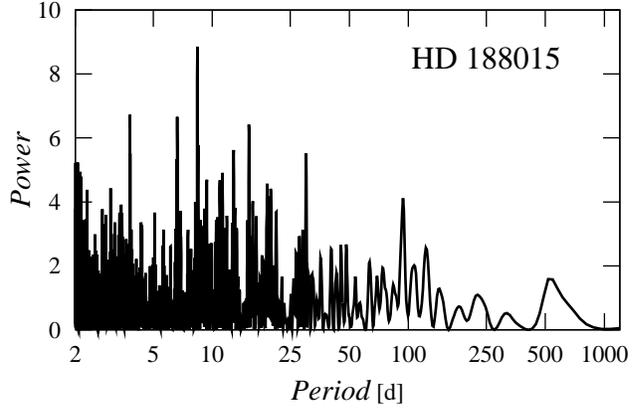}}
   \caption{
   Lomb-Scargle periodogram of the residuals of the long-period signal
   visible in the RV data of \starc{}.
}
\label{fig:fig15}%
\end{figure}

The results of the hybrid search perfectly agree with the LS periodogram
(Fig.~\ref{fig:fig15}) of the signal after removal of the contribution from the
outer planet. All the relevant periods can be found in the periodogram too. This
assures us that the GA search is robust in locating  the best-fit solution even
in the case when $\Chi$ has several equally deep local minima. Additionally, we
did an independent gradient-based scan over $P_{\idm{c}}$ and that test also
confirmed the excellent performance of the GAs.  

\begin{figure}
   \centering
   \hbox{
   \hbox{\includegraphics[width=3.5in]{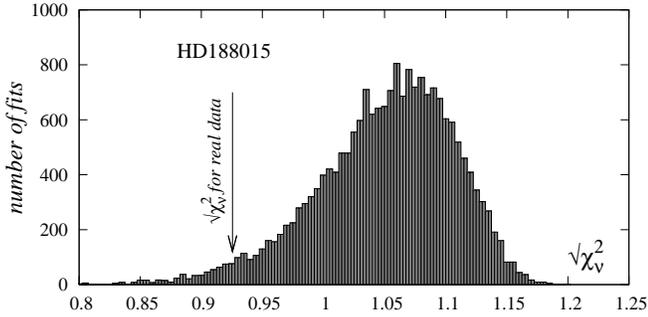}}
   }
   \caption{
   A histogram of $\Chi$ for Keplerian fits to 20,000 sets of scrambled RV
   residuals of the synthetic signal of the outer planet in \starc{}  (see
   Table~1). The position of the best-fit to the real data is marked with an arrow.
   }
\label{fig:fig16}%
\end{figure}

The test of  scrambled velocities which is illustrated in Fig.~\ref{fig:fig16}
helps us to explain the trouble of finding a clear, secondly  dominant
short-period in the data. The histogram of $\Chi$ has the maximum about of  1.1
and $\Chi\simeq 0.93$ of the real residuals  is not well separated from this
maximum.  The estimated $p_{\idm{H}} \simeq 0.03$ is  significant and  indicates
that the residual signal is rather noisy. Yet the best 2-planet fit improves
significantly the 1-planet solution.

We are warned by the referee (and we quote him again here) that the  jitter
estimate by \cite{Marcy2005} comes from the study of jitter levels in the
Carnegie planet search sample by \cite{Wright2005}. In that article, it turns
out that at the activity level of \starc{} corresponds to an average jitter of
the quoted 4~m/s, but with a wide dispersion from case to case, and that levels
of 6-7 m/s are not at all uncommon. In fact, for low values of the Calcium index
there is little activity-jitter correlation in \cite{Wright2005}, and any values
between 1 and 10~m/s are almost equally likely. This  may invalidate our
argument about the excess scatter and the value of $\Chi$ in the 2-planet 
model. Nevertheless, what we can do is to adopt an uniform jitter  variance for
all measurements. Then a larger jitter will only "flatten" $\Chi$, keeping its
minimum at similar position in the parameters space. The results for \starb{}
may be used as an argument supporting the planetary hypothesis ---in that case
the 2-planet fit also has an rms  about of 1~m/s smaller than the joint
uncertainty. Yet we encountered a similar problem when analyzing the next
system,  around \stard{}, described below.

\section{\stard}
%

\begin{figure*}
   \centering
   \hbox{
      \hbox{\includegraphics[width=3.4in]{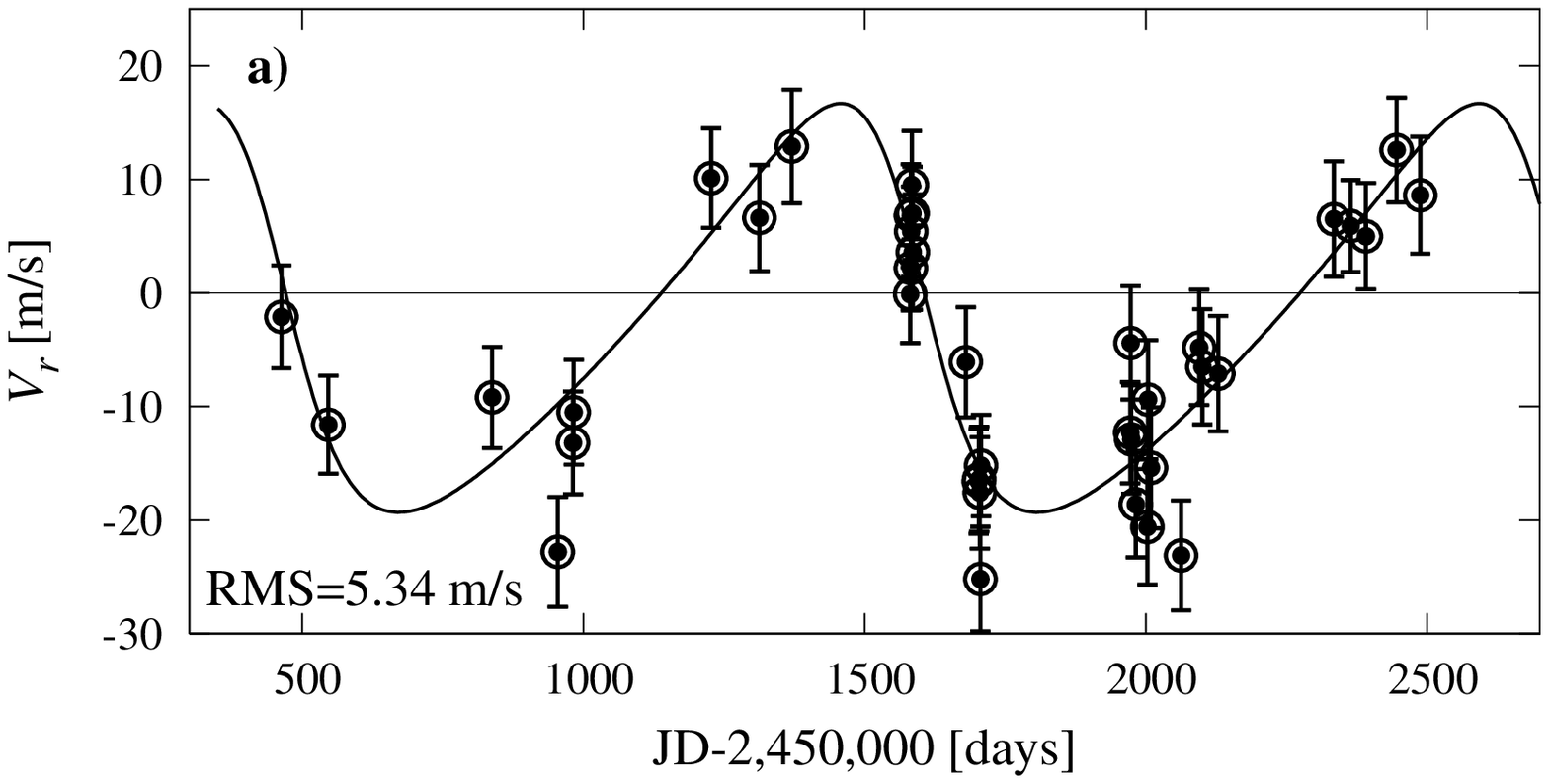}}
      \hbox{\includegraphics[width=3.4in]{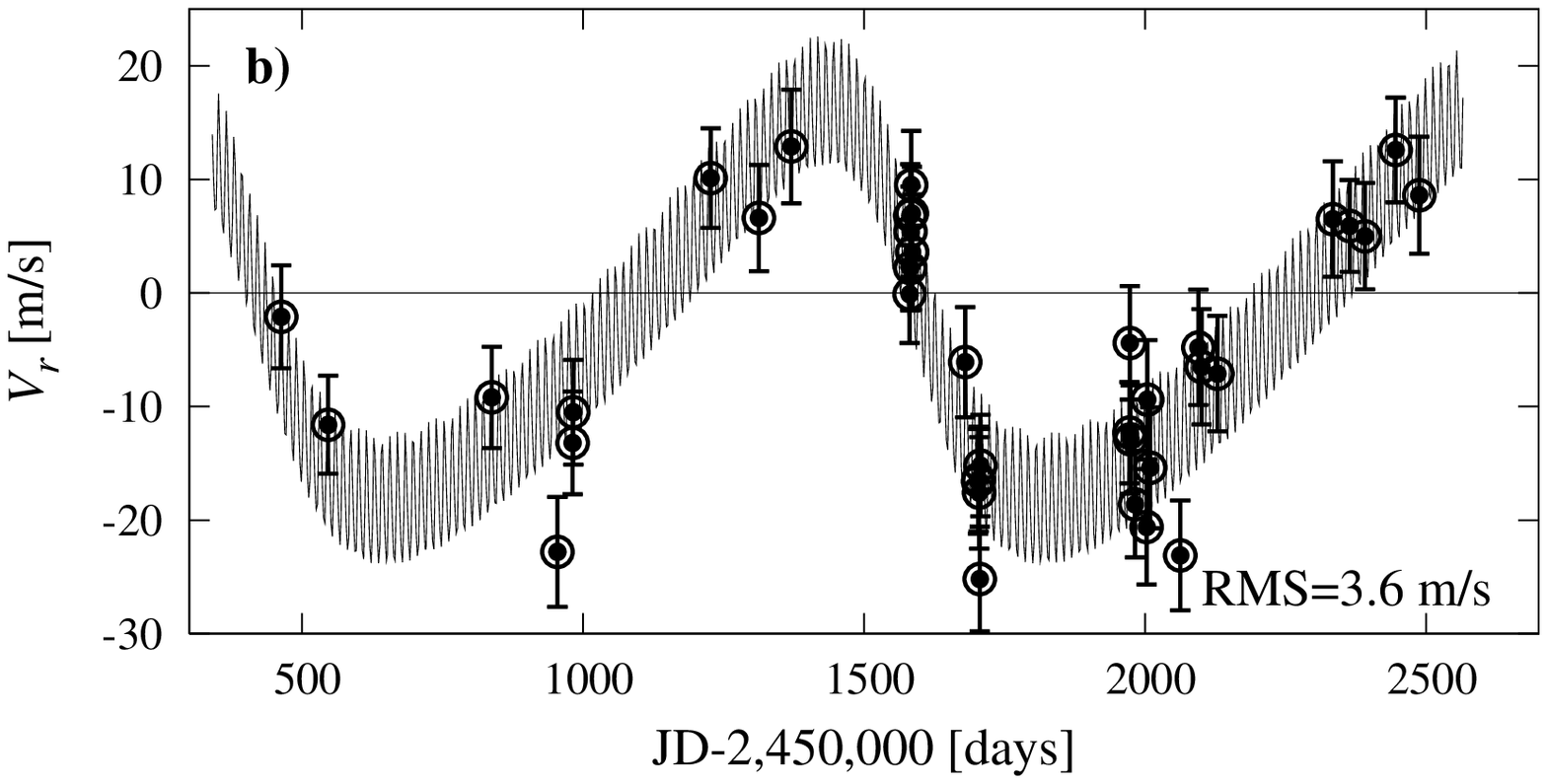}}
   }
   \hbox{
      \hbox{\includegraphics[width=3.4in]{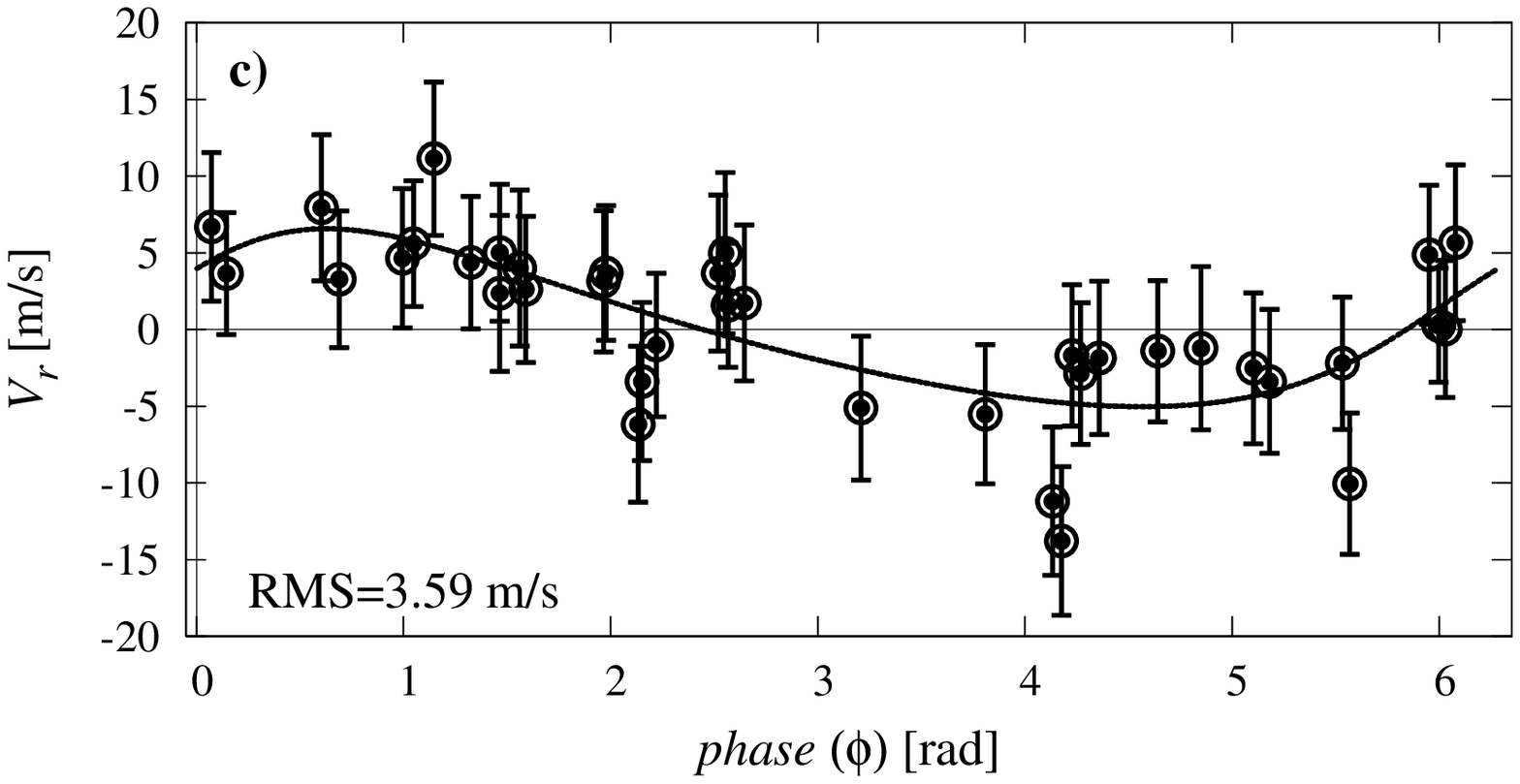}}
      \hbox{\includegraphics[width=3.4in]{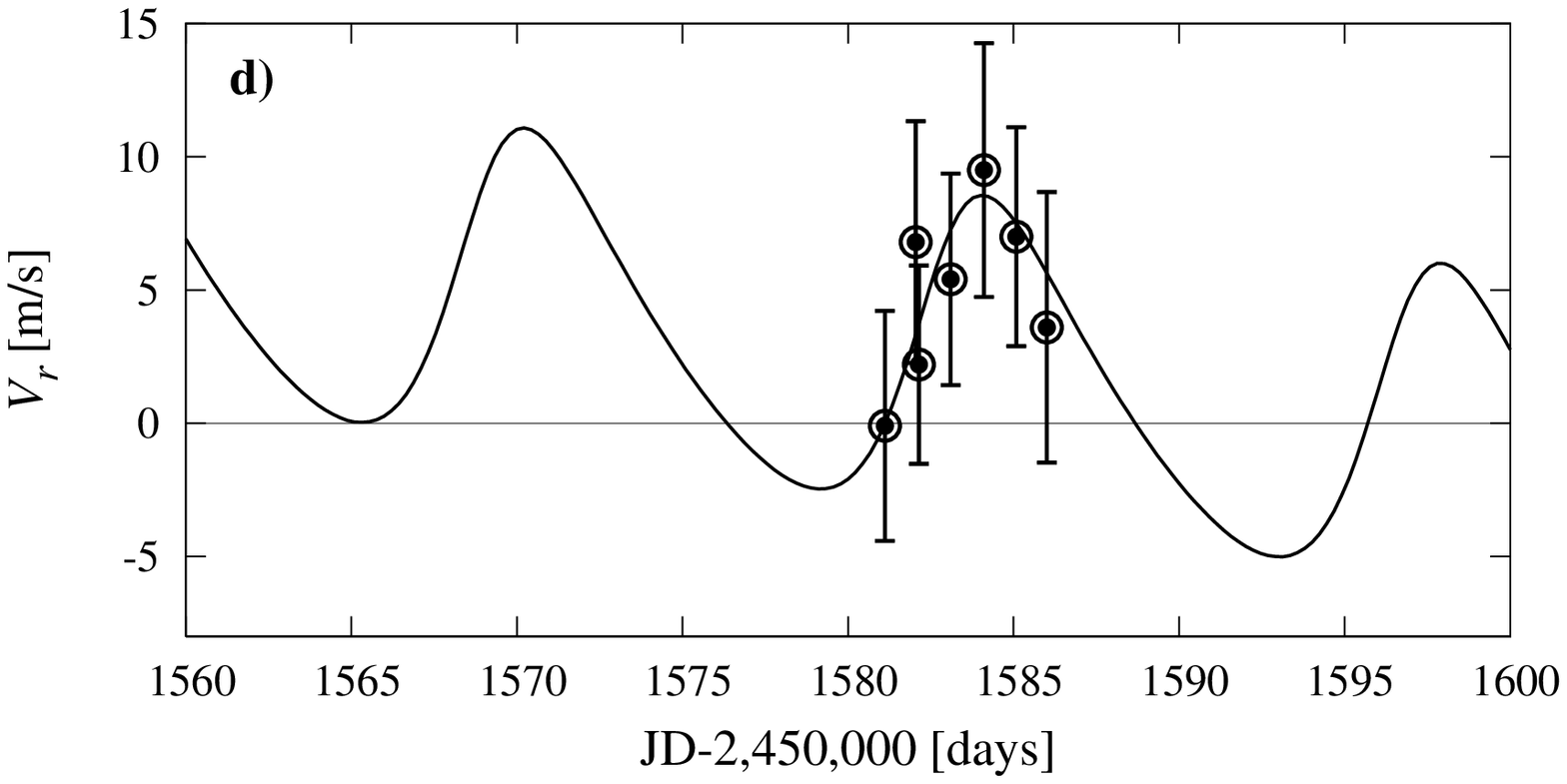}}
   }
   \caption{
    The best-fit solutions to the \stard{} RV data published in
\cite{Butler2003}. {\bf a}) Synthetic signal of 1-planet model. {\bf b})
2-Kepler model is consistent with the measurement error,  $\Chi\simeq 0.775$, an
rms $\simeq 3.6$~m/s.  {\bf c}) Period-phased  RV signal of the inner companion.
{\bf d}) A close-up of the synthetic signal of the best-fit 2-planet
configuration (see Table~1).
    }
\label{fig:fig17}
\end{figure*}

\begin{figure*}
 \hspace*{0.1cm}\centerline{
         \hbox{\includegraphics[]{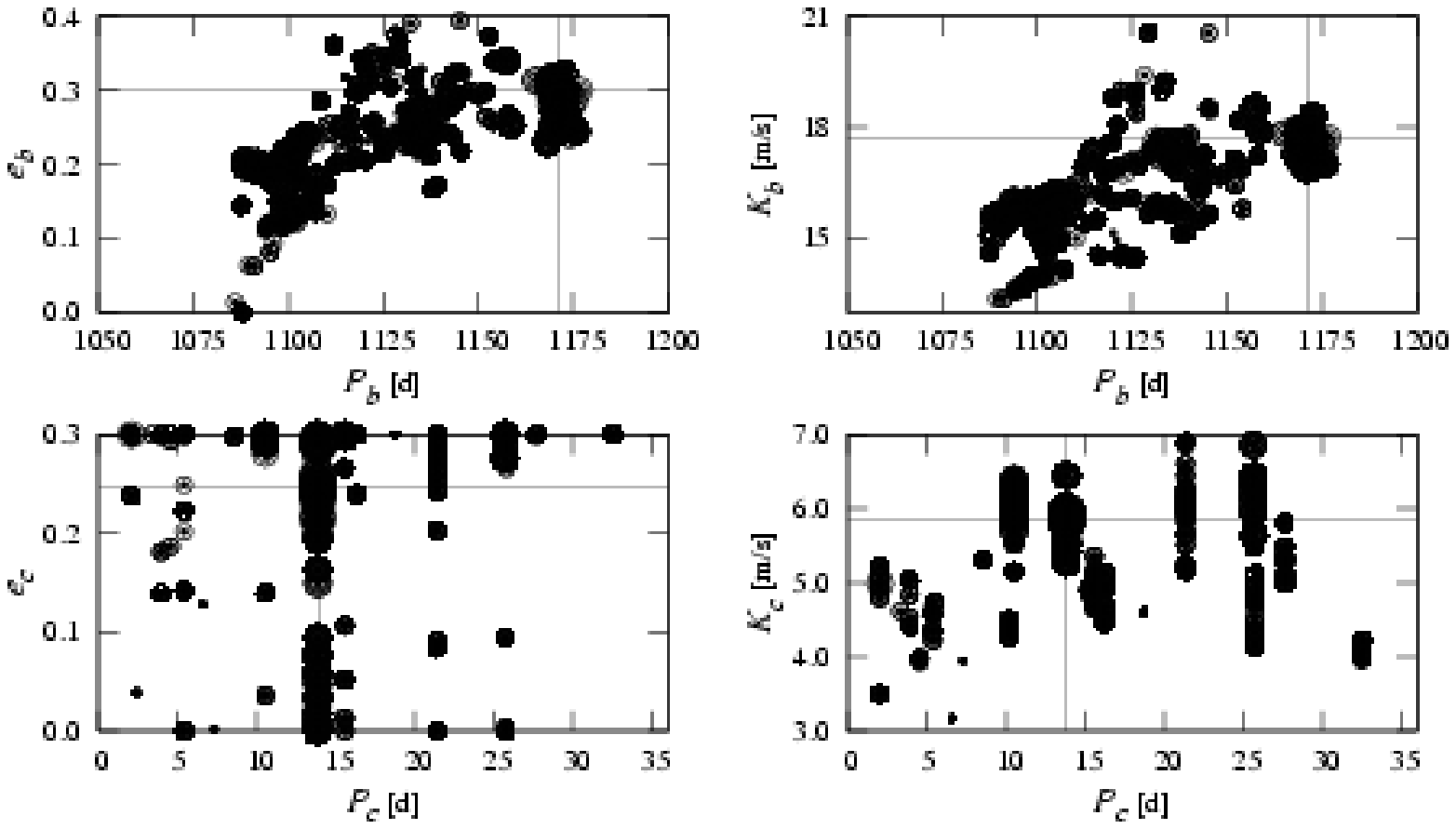}}}
 \caption{
 The parameters of the best-fits  to the 2-planet model of the \stard{} RV
 data  projected onto  the ($P,e$)- and ($P,K$)-plane.  Upper plots are for the
 outer planet, bottom plots are for the inner companion. The values of  $\Chi$
 of the best-fit solutions are marked by the size of symbols. Largest circle is
 for $\Chi$ equal to 0.775 (the best-fit solution, Table~1); smaller symbols are
 for 1$\sigma$ solutions with $\Chi \in [0.775,0.83)$,  2$\sigma$ solutions with
 $\Chi \in [0.83,0.915)$ and 3$\sigma$ solutions with $\Chi \in [0.915,1.02)$
 (smallest, filled circles), respectively.    
}
\label{fig:fig18}
\end{figure*}

\begin{figure}
   \centering
   \hbox{\includegraphics[]{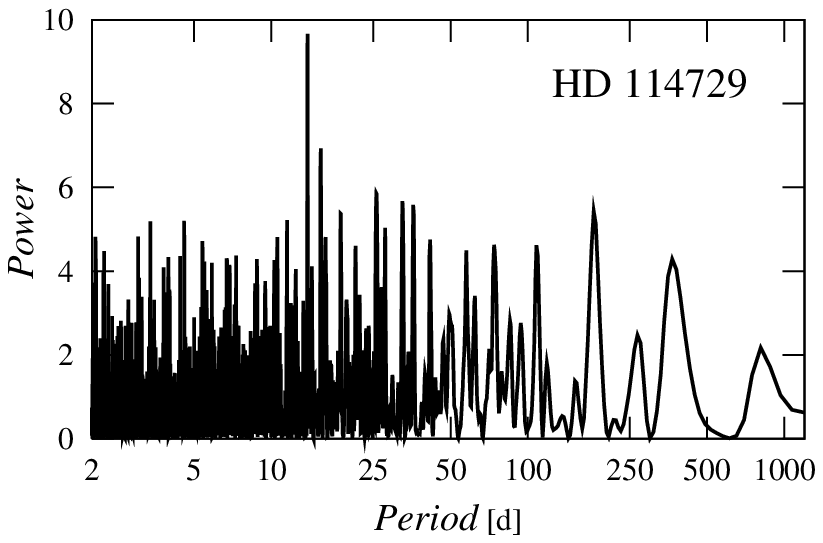}}
   \caption{
     Lomb-Scargle periodogram of the residuals of the long-period singal  in the
   \stard{} data.
}
\label{fig:fig19}
\end{figure}

Similarly to most of the stars monitored in precision RV surveys, the  \stard{} 
is a Sun-like star of G3V spectral type (G0V by Hipparcos).  Its RV data
indicate a Jupiter-like companion with minimal  mass 0.84~$m_{\idm{J}}$,
orbital  period $\simeq 3.1$~yr and eccentricity $\simeq 0.32$
\citep{Butler2003}. An rms of the single Keplerian fit to the data is 5.34~m/s.
It is larger than the mean of internal errors, $\simeq 3.6$~m/s.  The expected
jitter  is 4~m/s \citep{Butler2003}. The mean  measurement error scaled in
quadrature with the jitter variance  is comparable with the rms. Nevertheless,
having in mind a significant scatter of the measurements about some parts of the
synthetic 1-planet RV curve (see Fig.~\ref{fig:fig17}a), and the presence of the
distant giant planet, we assumed (as in the previous case) that there exist an
another one, inner planet in the system in short-period orbit, [2,136]~d. To
test this hypothesis, we did the same analysis of the RV data as described in
previous sections.  The results are illustrated in Fig.~\ref{fig:fig17}a,b,c,d
which is for the graphical illustration of the best-fit parameters.
Figure~\ref{fig:fig18} is for the Lomb-Scargle periodogram of the residual
signal of the long-period companion.  Figure~\ref{fig:fig19}  illustrates  the
shape of $\Chi$ about the best-fit and the distribution of the best fits
obtained in the hybrid search.  The orbital parameters of the best fit are given
in Table~1. Clearly, the results of the hybrid search agree perfectly with the
LS periodogram---the dominant  peak about of 7~d is clearly visible in the
$(P_{\idm{c}},e_{\idm{c}})$-plane as the deepest minimum of $\Chi$ in the region
of short periods. The best-fit eccentricity of the inner planet is found  close
to the assumed limit of 0.3. However, in a few tests with smaller jitter  ($\sim
3$~m/s) we obtained better bounded minimum of $\Chi$ and smaller best-fit 
$e_{\idm{c}} \simeq 0.2$, so this element is barely constrained by the
RV data.

An inspection of the LS periodogram shows some other relatively  strong peaks,
for instance about of 180~d, which most likely are aliasing  with the period of
$13.844$~d. 

\begin{figure}
   \centering
   \hbox{
   \hbox{\includegraphics[width=3.5in]{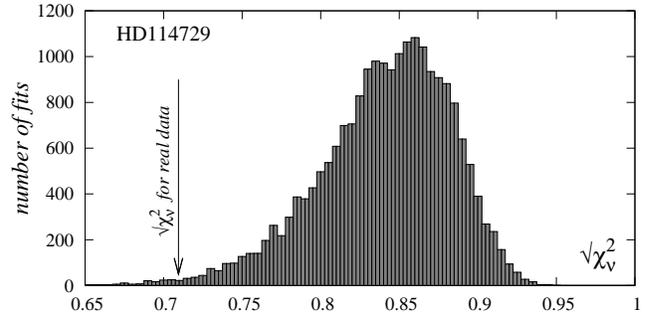}}
   }
   \caption{
   A histogram of $\Chi$ for Keplerian fits to 21,000 sets of scrambled
   RV residuals of the synthetic signal of the outer planet in the \stard{}
   planetary system (Table~1). The position of the best-fit to the real data is
   marked with an arrow. 
   }
\label{fig:fig20}
\end{figure}

Figure~\ref{fig:fig20} is  for the histogram of $\Chi$  obtained for scrambled
residuals of the outer planet's signal.  The probability that the residual
signal of the outer companion  has a random origin is very small,
$p_{\idm{H}}\simeq 0.003$.  The $\Chi$ for the  best-fit solution has well
defined minimum about of 0.77. Its rms of $\sim 3.6$~m/s is very close to the
mean of the internal errors. Assuming that  the mass of the host star is equal
to $0.93~M_{\sun}$, the  derived minimal mass of the inner planet is about of 
$0.065 m_{\idm{J}}$ and its semi-major axis $\simeq 0.11$~AU.  Thus it would
correspond to a hot Neptune-like planet.

\section{\stare{}}
%
The last example which we examine is \stare{}. This star is observed by
\cite{Mayor2004} who published 30 measurements covering 1690~d. In this case, an
rms of 5.7~m/s of the 1-planet fit is  comparable with the mean of measurement
errors, nevertheless our attention draws an irregular scatter of the data about
the synthetic RV curve, Fig.~\ref{fig:fig22}a. The discovery team announced a
Jupiter-like planet  of minimal mass 1.21~$m_{\idm{J}}$ in $\simeq 528$~d orbit.
The host star is very active ($R_{\idm{HK}}'=-4.4$). Following \cite{Wright2005}
we estimate its jitter $\sim 9$~m/s.

\begin{figure*}
   \hspace*{0.1cm}\centerline{
         \hbox{\includegraphics[]{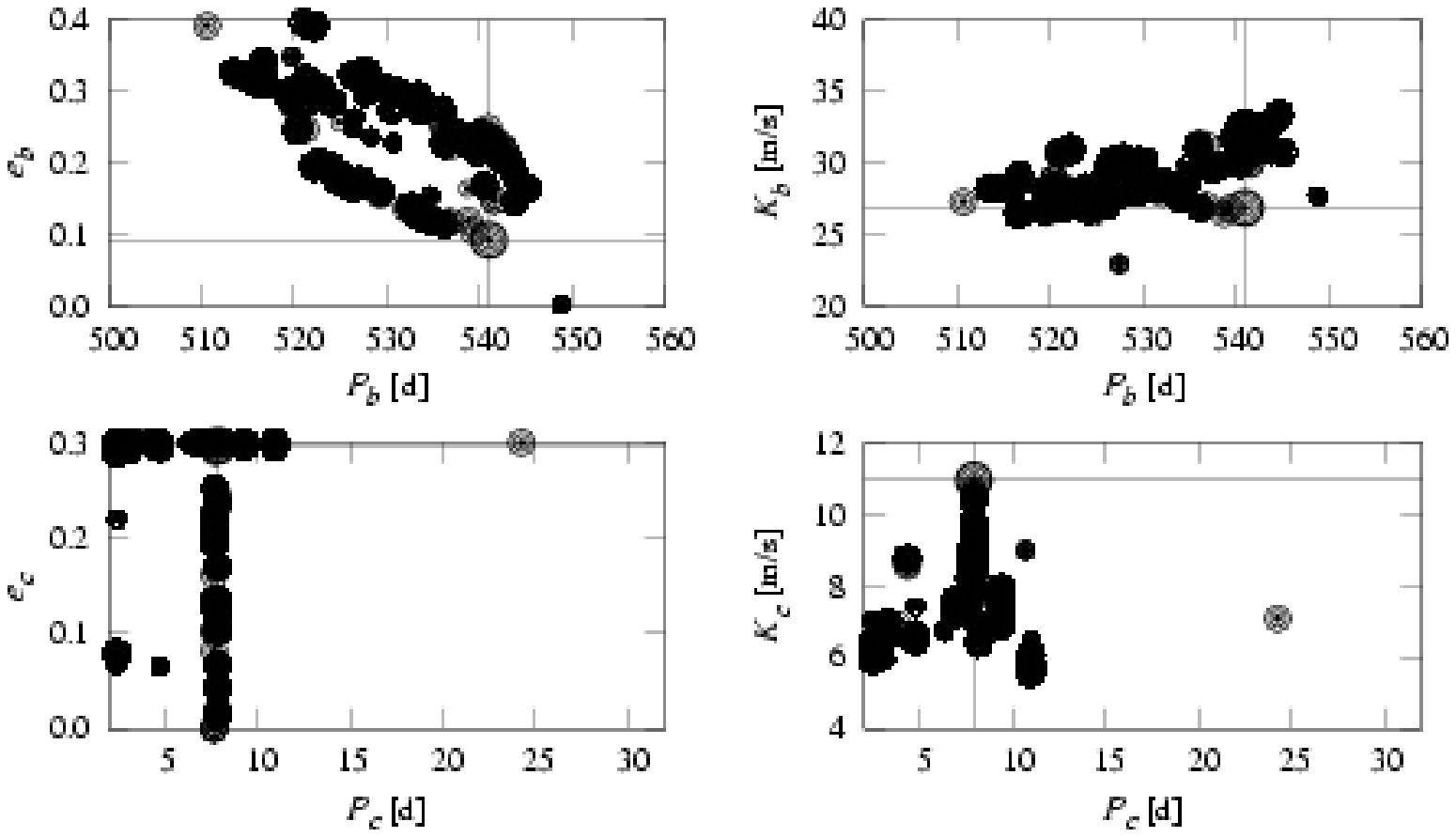}}}
   \caption{
   The projections of best-fit parameters to the 2-planet model of the \stare{}
RV data  onto the ($P,e$)- and ($P,K$)-plane.  Upper plots are for the outer
planet, bottom plots are for the inner companion. The values of  $\Chi$ of the
best-fit solutions are marked by the size of symbols. Largest circle is for
$\Chi \simeq 0.505$ (the best-fit solution, Table~1); smaller symbols are for
1$\sigma$ solutions with $\Chi \in [0.505,0.619)$,  2$\sigma$ solutions with
$\Chi \in [0.62,0.773)$ and 3$\sigma$ solutions with $\Chi \in [0.773,0.95)$
(smallest, filled circles), respectively.    }
\label{fig:fig21}
\end{figure*}

\begin{figure*}
   \centering
   \hbox{
      \hbox{\includegraphics[width=3.4in]{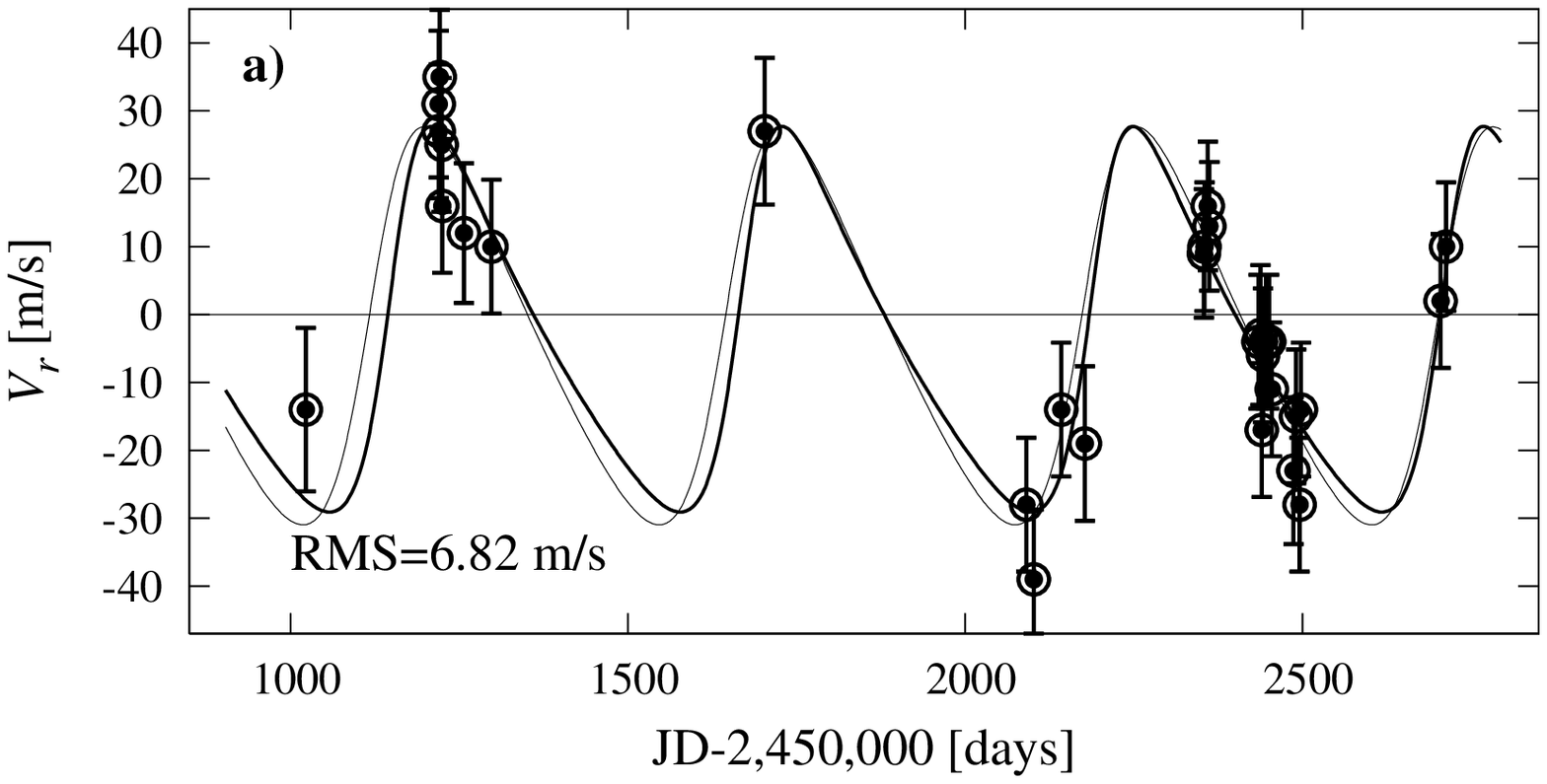}}
      \hbox{\includegraphics[width=3.4in]{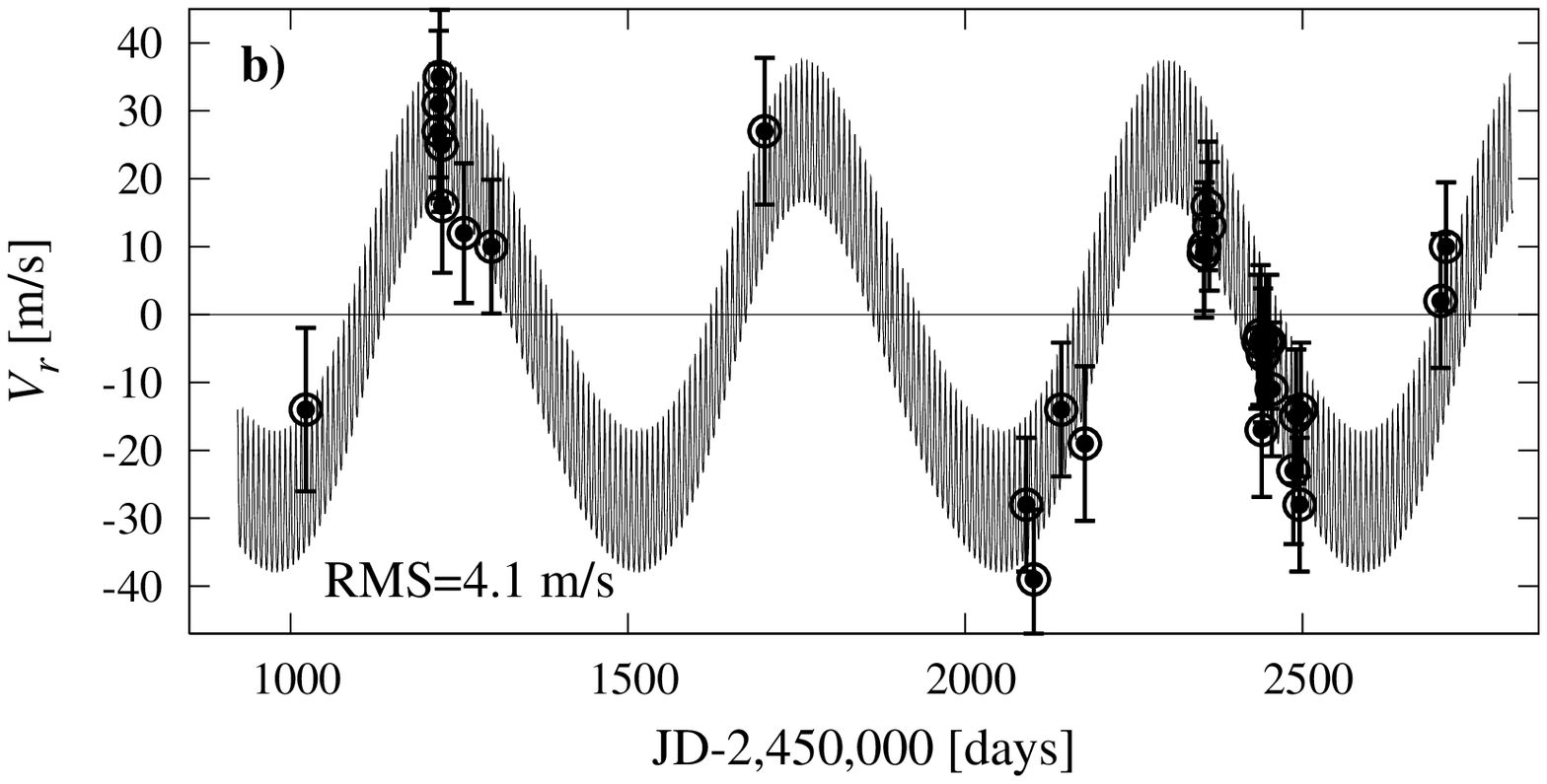}}
   }
   \hbox{
      \hbox{\includegraphics[width=3.4in]{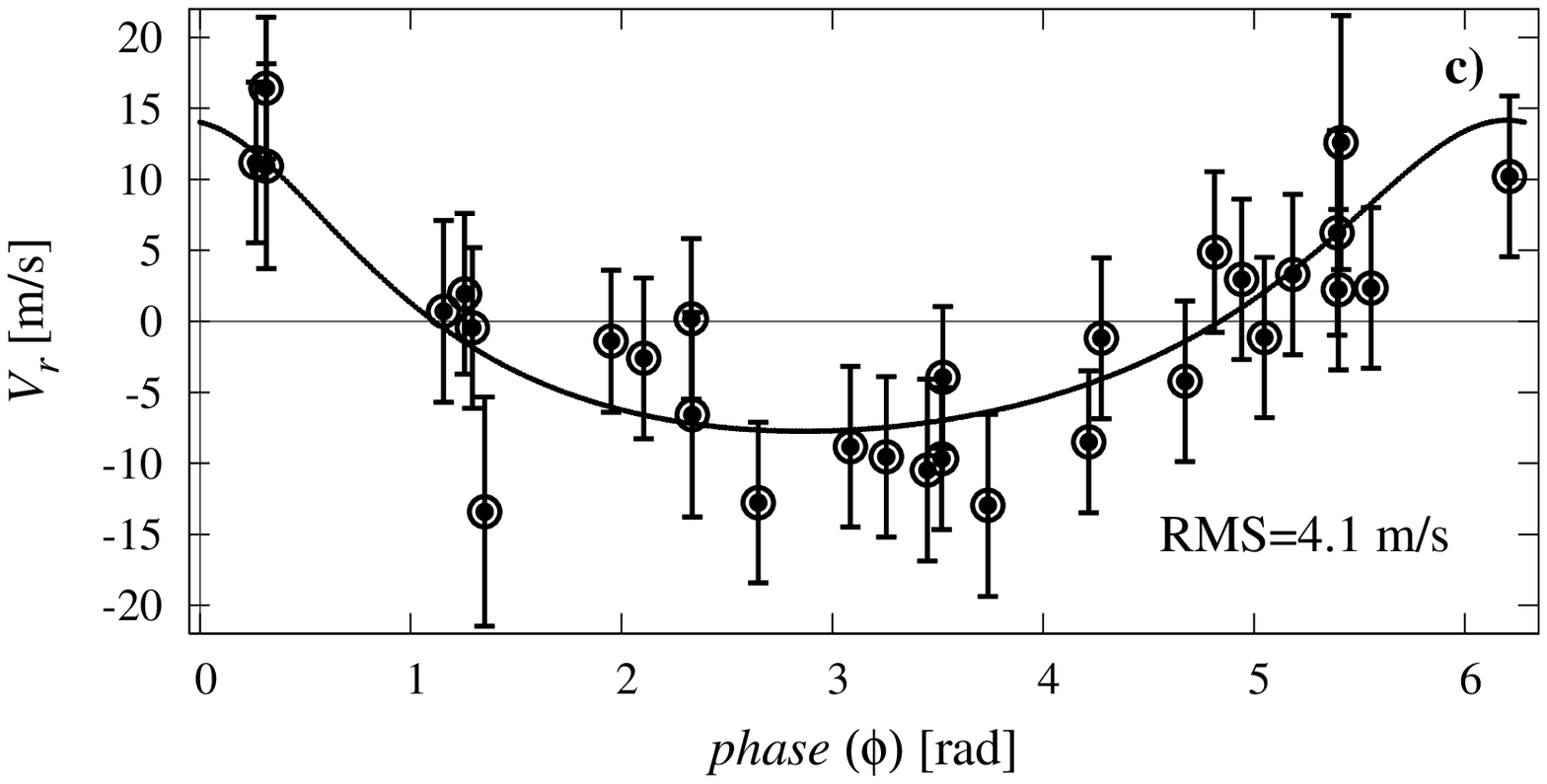}}
      \hbox{\includegraphics[width=3.4in]{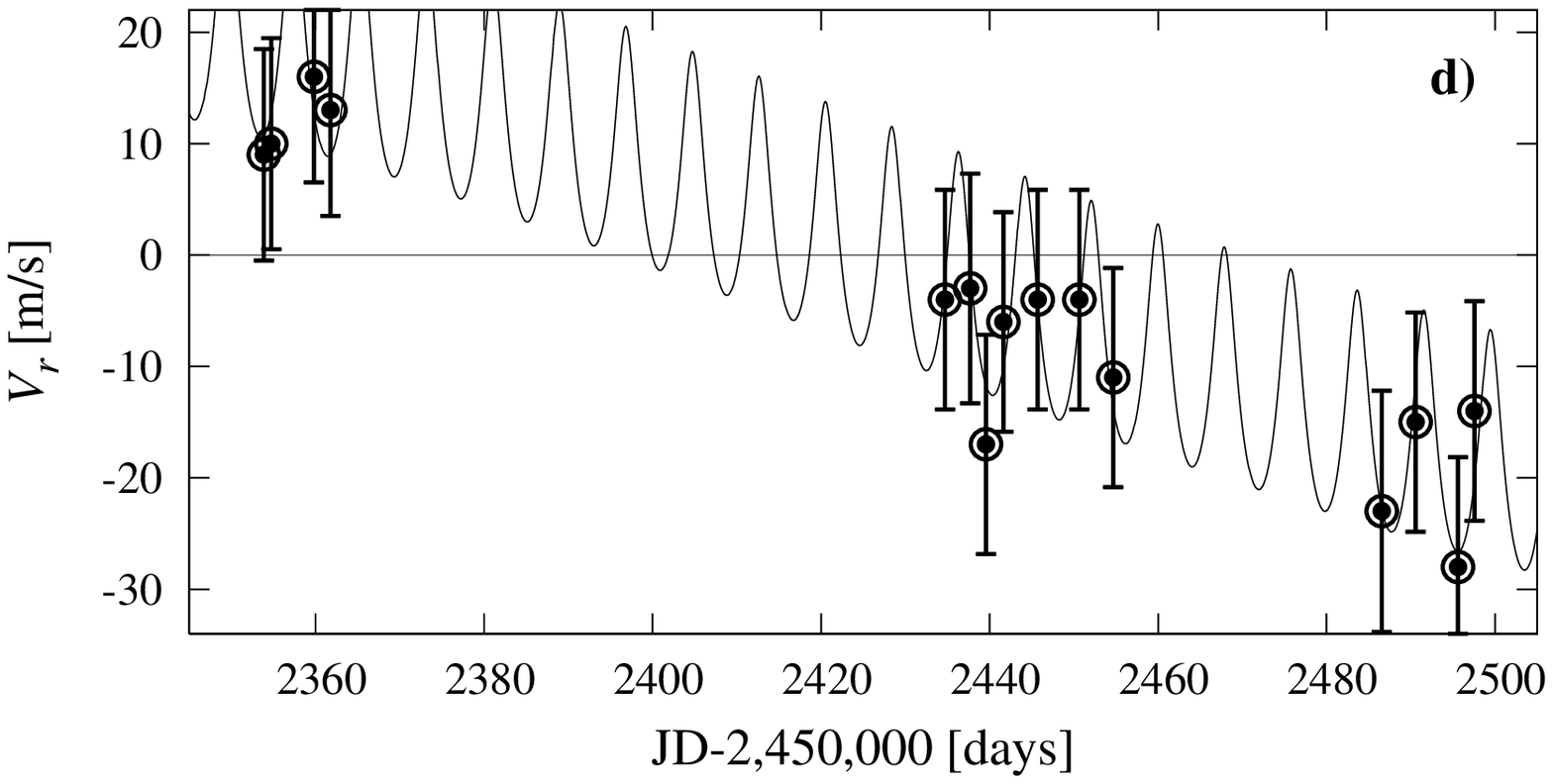}}
   }
   \caption{
    The best-fit solutions to the RV data published in \citep{Mayor2004} for
    \stare. {\bf a}) 1-planet model. The thick line is for the fit found with
    GA, assuming $\sigma_{j}=9$~m/s; its parameters 
    $(K,P,e,\omega,T_{\idm{p}}-T_0)$ are:     
    $(28.40~\mbox{m/s},520,49~\mbox{d},
    0.346,288^{\circ}.57,3125.302~\mbox{d})$, $T_0$=JD~2,450,000, $\Chi=0.737$,
    an rms=$6.83$~m/s. The thin line is for the fit {\em without accounting} for
    the jitter; its parameters are almost the same as in \citep{Mayor2004}:
    ($29.312~\mbox{m/s},528.112~\mbox{d},
    0.260,282^{\circ}.453,2214.94~\mbox{d}$), $\Chi=1.60$, an rms $=7.26$~m/s.
    {\bf b}) 2-Kepler model is consistent with the measurement errors,  $\Chi
    \simeq 0.505$, an rms$\simeq 4.1$~m/s.  {\bf c}) Period-phased  RV signal of
    the inner companion. {\bf d}) A close-up of the synthetic signal of the
    best-fit 2-planet configuration, Table~1.
    }
\label{fig:fig22}
\end{figure*}

We searched for the second planet in short-period orbit, [2,136]~d.  The results
of the hybrid search obtained for 2-planet fits are illustrated in
Fig.~\ref{fig:fig21} for the statistics of the  gathered best-fits and in
Fig.~\ref{fig:fig22}a,b,c,d for the graphical illustration of the  best-fit
solution. These results remind us  the \starb{} case.  The $\Chi$ of the
best-fit  has well defined minimum at $P_{\idm{c}}\simeq 7.894$~d. Its rms
$\simeq 4.1$~m/s means a significant improvement of the  single-planet fit. This
is seen in Fig.~\ref{fig:fig22}b,d; apparently the synthetic RV curve is ideally
close to the measurements. A relatively large $K_{\idm{c}}\sim 11$~m/s and  mass
of the host star $1.11~M_{\sun}$ imply the minimal mass of the inner planet 
$0.115~m_{\idm{J}}$ and semi-major axis about of $0.08$~AU. 

\begin{figure}
   \centering
   \hbox{\includegraphics[]{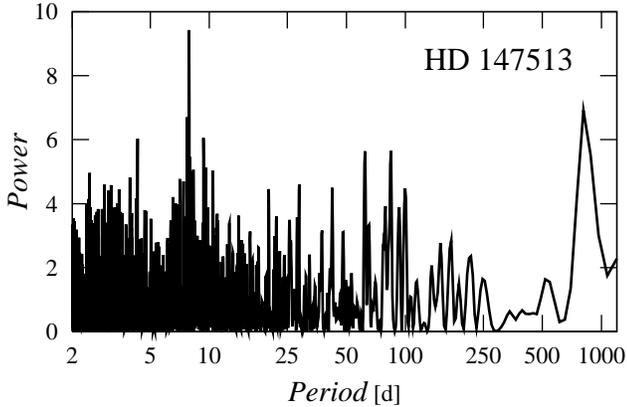}}
   \caption{
Lomb-Scargle periodogram of the residuals of  the long-period
   signal visible in the RV data of \stare{}.
}
\label{fig:fig23}%
\end{figure}

\begin{figure}
   \centering
   \hbox{
   \hbox{\includegraphics[width=3.5in]{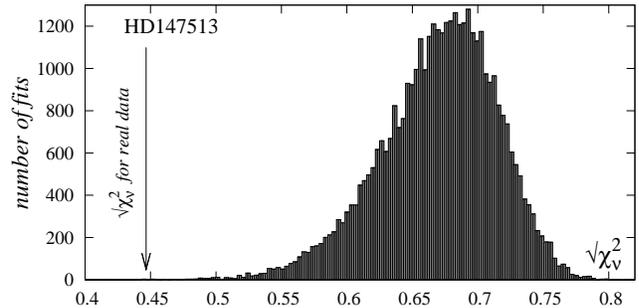}}
   }
   \caption{
   A histogram of $\Chi$ for Keplerian fits to 30,000 sets of scrambled RV
   residuals of the synthetic signal of the outer planet in the \stare{}
   planetary system (Table~1). The position of the best-fit to the real data is
   marked with an arrow.  }
\label{fig:fig24}
\end{figure}

The LS periodogram, Fig.~\ref{fig:fig23},  reveals features which recall
us the first system examined in this paper, about \stara{}. The short period of
$\simeq 7.9$~d  seems to be aliased with a much longer period of $810.8$~d. 
Figure~\ref{fig:fig24} is for the histogram of $\Chi$  computed for the scrambled
residuals. The probability of a random character of the residual signal  is
negligible.  At this time, such a characteristic is almost the same as for the
\starb{} system; in that case  the 2-planet solution also produces very small
rms which is less than the joint measurement uncertainty.

Still, mostly due to the activity of the star,  the periodicity of the RV
variations can be explained by jitter.  According to \cite{Mayor2004},  \stare{}
is a young G3/5V dwarf, only 0.3Gyr old. Its spectrum reveals a strong  emission
in the Ca~II~H line. Although  the estimated short rotation period of the host
star, about of 4.7~d \citep{Mayor2004},  differs from the secondary period
visible  in the RV signal (about of 7.9~d), it is suspiciously close to this
approximation. However, such coincidence of periods still does not exclude a
planetary explanation \citep{Santos2000,Santos2003} because the activity can be
induced or amplified by relatively massive and close planetary companion.

\begin{table*}
\label{tab:tab1}
\caption{
Primary parameters of the model Eq.~1  and inferred elements of the best-fits
found in this paper, $T_0$ is JD~2,450,000. See the text and figures for error
estimates.
}
\centering
\begin{tabular}{lcccccccccc}
\hline
\hline
&\multicolumn{2}{c}{\starc} &\multicolumn{2}{c}{\stard}
&\multicolumn{2}{c}{\stare} &\multicolumn{2}{c}{\starb}
&\multicolumn{2}{c}{\stara}
\\
\hline
Parameter \hspace{1em}  
& \ \  {\bf c} \ \  & \ \ {\bf b} \ \ 
& \ \  {\bf c} \ \  & \ \ {\bf b} \ \ 
& \ \  {\bf c} \ \  & \ \ {\bf b} \ \ 
& \ \  {\bf c} \ \  & \ \ {\bf b} \ \ 
& \ \  {\bf c} \ \  & \ \ {\bf b} \ \ 
\\
\hline
$M_{\star}$ [$M_{\sun}$]
                        & \multicolumn{2}{c}{1.08}  & 
			  \multicolumn{2}{c}{0.93}  & 
			  \multicolumn{2}{c}{1.11}  &
			  \multicolumn{2}{c}{0.95}  &
			  \multicolumn{2}{c}{0.96}  
\\			
$m \sin i$ [M$_{\idm{J}}$]  
			&   0.076  &    1.81    
			&   0.066  &    0.88      
                        &   0.115  &    1.15
                        &   0.064  &    1.62
                        &   0.121  &    0.46        
\\
$a$ [AU] 		&   0.08   &    1.22   
			&   0.11   &    2.12
                        &   0.08   &    1.35
                        &   0.13   &    3.93      
                        &   0.11   &    0.49            
  \\
$P$ [d]  		&   8.389  &  476.23
                        &  13.849  & 1171.19
                        &   7.894  &  541.00
                        &  17.104  & 2908.32
			&  14.496  &  129.51
\\
$K$ [m/s] 		&   7.21   &   44.75
			&   5.87   &   17.71
                        &  10.95   &   26.85
                        &   5.20   &   24.90       
			&  10.47   &   19.08
\\
$e$     		&   0.23   &    0.24 
			&   0.25   &    0.30
                        &   0.30   &    0.09
                        &   0.21   &    0.37
			&   0.43   &    0.26
\\
$\omega$ [deg] 		& 103.9    &  242.4  
			& 301.1    &   66.3 
                        &   9.1    &  316.8 
                        &  10.1    &   10.8       
			&  27.6    &  130.9
\\
$T_{\idm{p}}$ [JD-$T_0$]
			&  859.67  &  2270.10
			& 2496.77  &  1558.98      
                        & 1047.23  &  2248.44
                        & 2370.38  &  3517.82
			& 2815.87  &  2957.74
\\
$\Chi$  		& \multicolumn{2}{c}{1.021} & 
		 	  \multicolumn{2}{c}{0.775} &
		 	  \multicolumn{2}{c}{0.505} &
		 	  \multicolumn{2}{c}{0.680} &
		 	  \multicolumn{2}{c}{0.581} 
\\
$\sigma{\idm{j}}$~[m/s] & \multicolumn{2}{c}{4.0}  & 
			  \multicolumn{2}{c}{4.0}  & 
			  \multicolumn{2}{c}{9.0}  &
			  \multicolumn{2}{c}{3.1}  &
			  \multicolumn{2}{c}{3.0}  
\\
$\overline{\sigma}^2$~[m/s] 
                        & \multicolumn{2}{c}{4.1}  & 
			  \multicolumn{2}{c}{3.5}  & 
			  \multicolumn{2}{c}{4.4}  &
			  \multicolumn{2}{c}{2.2}  &
			  \multicolumn{2}{c}{5.4}  
\\
$(\sigma{\idm{j}}^2+\overline{\sigma}^2)^{1/2}$~[m/s] 
                        & \multicolumn{2}{c}{5.7}  & 
			  \multicolumn{2}{c}{5.3}  & 
			  \multicolumn{2}{c}{10}  &
			  \multicolumn{2}{c}{3.8}  &
			  \multicolumn{2}{c}{6.2}  
\\
rms~[m/s] 		& \multicolumn{2}{c}{4.93}  & 
			  \multicolumn{2}{c}{3.59}  & 
			  \multicolumn{2}{c}{4.07}  &
			  \multicolumn{2}{c}{2.97}  &
			  \multicolumn{2}{c}{2.82}  
\\
$V_0$ [m/s] 		& \multicolumn{2}{c}{-17.45}  &
			  \multicolumn{2}{c}{-3.22} &
			  \multicolumn{2}{c}{-4.85} &
			  \multicolumn{2}{c}{-4.81} &
			  \multicolumn{2}{c}{7.97} 
\\ 
\hline
\end{tabular}
\end{table*}

\section{Conclusions}
%

It is extremely difficult to interpret the RV measurements when their number is
limited and their time coverage is poor. We are looking for signals  which have
an amplitude comparable with noise. Although is already possible to pick up some
promising  extrasolar systems in which one planet is clearly visible and an another
putative body is at the detection limit, a claim that the residual signal has a
planetary origin is always very risky. To confirm the existence of the putative
hot-Neptune candidates, a deep astrophysical analysis is necessary
\citep{Mayor1995,Santos2003,Butler2004}, by far beyond our technical abilities.
That includes many additional spectroscopic and photometric observations.
Nevertheless,  the dynamical environment of the putative new planetary
companions, several recent discoveries of Neptune-like planets and  cosmogonic
theories supporting  their origin, permit us to state  and investigate a
hypothesis of the planetary nature of the RV variability.

Our results may be useful in some aspects. Even if the residual periodicities
are of pure stellar origin, accounting them for may improve the  single planet
fits, because it makes it possible to remove a  contribution of not-random RV
signal from the data. When we deal with a small number of measurements, the
short-term contribution of jitter does not average out and it alters the
long-period solution. Yet the stellar variations may have origin in
planetary-induced activity.  This effect has been analyzed by
\cite{Santos2000,Santos2003}.  An elimination of the planetary hypothesis also
requires additional observations and analysis.  

Basically, we use an independent, FFT-free method of the  global analysis of the
RV data and early detection of plausible planets. It can be used for checking
the results of 1-dim periodogram-based analysis, extending it into the
multidimensional space of the orbital elements.  If some of our results will be
confirmed,  a similar approach can be used to detect  smaller bodies in
short-period orbits in the other,  already discovered extrasolar systems. Such
studies can be useful for planning future observing sessions.

We did a similar but very preliminary analysis of the  RV observations of a few
other stars hosting single Jupiter-like planets.  For instance, we found a
short-term variability of the RV, about of 7~d in the HD~19994 system
\citep{Mayor2004}. The second companion would have $K\sim 9$~m/s, larger than 
the mean uncertainty of the measurements, about of 6.6~m/s.  In the same paper,
we found the RV data of HD~121504 and a curious RV variability which may be  well
modeled by  a new 2-planet system close to the 6:1 
mean motion resonance. Likely, the data
collected and published by Doppler planet  searches already hide many
interesting and unusual planetary configurations.

\begin{acknowledgements}
We thank the anonymous referee for comments and  invaluable suggestions that
improved the manuscript. We deeply acknowledge the
pioniering work  and  publishing of the precision RV data by planet search
teams.  K.G. thanks Jean
Schneider for the inspiration to study the \stara{} system. This work is
supported by Polish Ministry of Sciences and Information  Society Technologies, 
Grant No. 1P03D-021-29.
\end{acknowledgements}
\bibliographystyle{aa}
\bibliography{ms}
\end{document}